\documentclass[fleqn,usenatbib]{mnras}

\usepackage{newtxtext,newtxmath}
\usepackage[T1]{fontenc}

\DeclareRobustCommand{\VAN}[3]{#2}
\let\VANthebibliography\thebibliography
\def\thebibliography{\DeclareRobustCommand{\VAN}[3]{##3}\VANthebibliography}

\usepackage{graphicx}
\usepackage{amsmath}

\title[Strange quark matter as dark matter]{Strange quark matter as dark matter: 40 years later, a reappraisal}

\author[Francesco Di Clemente et al.]{
Francesco Di Clemente $^{1}$
,
Marco Casolino $^{2}$,
Alessandro Drago $^{1,3}$,
Massimiliano Lattanzi $^{1}$,
Claudia Ratti$^{4}$
\\
$^{1}$INFN, Sezione di Ferrara, via Saragat 1, I-44122 Ferrara, Italy\\
$^{2}$INFN and University of Rome, Tor Vergata, Italy\\
$^{3}$Department of Physics and Earth Science, University of Ferrara, via Saragat 1, I-44122 Ferrara, Italy\\
$^{4}$Department of Physics, University of Houston, Houston, TX 77204, USA
}

\date{Accepted XXX. Received YYY; in original form ZZZ}

\pubyear{2024}

\begin{document}
\label{firstpage}
\pagerange{\pageref{firstpage}--\pageref{lastpage}}
\maketitle

\begin{abstract}
Forty years ago Witten suggested that dark matter could be composed of macroscopic clusters of strange quark matter. This idea was very popular for several years, but it dropped out of fashion once lattice QCD calculations indicated that the confinement/deconfinement transition, at small baryonic chemical potential, is not first order, which seemed to be a crucial requirement in order to produce large clusters of quarks. 
Here we revisit the conditions under which strangelets can be produced in the Early Universe. We discuss the impact of an instability in the hadronic phase separating a low density, positive-strange-charge phase from a high-density phase with a negative strange charge. This second phase can rapidly stabilize by forming color-superconducting gaps. The strangelets then undergo partial evaporation. In this way, we obtain distributions of their sizes in agreement with the observational constraints and we discuss the many astrophysical and cosmological implications of these objects. Finally, we examine the most promising techniques to detect this type of strangelets. We also show that strangelets can exist with masses $\lesssim10^{17} \mathrm g$, while primordial black holes are ruled out in that mass range, allowing us to distinguish between these two dark matter candidates.
\end{abstract}

\begin{keywords}
cosmology: dark matter - cosmology: early Universe - dense matter.
\end{keywords}

\section{Introduction}

The nature of dark matter remains elusive, also for its derived distribution in galaxies \citep{Salucci:2018hqu}. One possibility is that dark matter is macroscopic, made of smaller
components that can be described within the standard model \citep{Burdin:2014xma,Jacobs:2014yca}. About forty years ago, it was suggested that dark matter could be made of strangelets, clusters of up, down, and strange quarks \citep{Witten:1984rs} formed at the time of the QCD phase transition in the Early Universe. The
important point is that dark matter made of strangelets can be effectively weakly interacting because it is extremely massive.
This possibility is based on the Bodmer-Witten (BW) hypothesis \citep{Bodmer:1971we,Witten:1984rs}, stating that clusters of strange quark matter (SQM) can be more stable than the most stable atomic nucleus, i.e. $^{56}$Fe. The reason why huge amounts of Fe surround us, while we struggle to find any trace of SQM, is that SQM needs a finite and not-too-small fraction of strange quarks in order to be stable. Since multiple, simultaneous weak decays are enormously suppressed, it is clear that the large amount of strangeness needed to satisfy the BW hypothesis can be obtained only under very specific conditions. To date, the only situations in which those conditions can be satisfied are the Early Universe and the core of massive stellar objects. 

It is important to note that the validity of the BW hypothesis does not imply that dark matter is made of SQM: it is possible that the hadronization process in the Early Universe does not allow the production of the large clusters of SQM that would constitute dark matter (the possibility that dark matter is made of small clusters of SQM is probably ruled out by observations, as we will discuss in the following). 

Still, the BW hypothesis could be true, and strange quark stars (QSs) made entirely of SQM could exist \citep{Haensel:1986,Weber:2004kj,Schaffner-Bielich:2004lxd,Madsen:1998uh}, since the mechanism of production of SQM in compact stars (which is based on the large chemical potential reached in those objects) is completely different from the one in the Early Universe. On the other hand, if dark matter is indeed composed of SQM, it is very unlikely that QSs do not exist, because many concrete astrophysical paths would lead to their production. This would still not imply that all compact stars are QSs! It is possible to show that, at least in the areas of the galaxies where dark matter is not so abundant, neutron stars (NSs) can exist without being transformed into QSs. This is of paramount importance, since it is unlikely that all compact stars are QSs and that NSs do not exist, see e.g. \citet{Watts:2006hk}. Indeed, in the last ten years, a scheme in which NSs and QSs coexist has been developed, the so-called two-family scenario \citep{Drago:2013fsa,DragoLavagnoHyperons2014,Drago:2015cea,Drago:2015dea,Wiktorowicz:2017swq}. In that scenario, the most massive objects are QSs; instead, NSs can have small radii, but their maximum mass does not reach 2$\mathrm{M}_\odot$. In addition, it has been proposed that QSs can be distinguished from NSs through their cooling behaviors \citep{Madsen1998} and r-mode instabilities in millisecond pulsars \citep{Madsen1999}.

The very first studies on the equation of state (EoS) of quark matter in compact stars were, in general, based on the MIT bag model \citep{Chodos1974} and discussed
strangeness production \citep{Freedman:1977gz}. Three major developments took place after those first pioneering works. 1) Chiral models became popular, based on many variations of the Nambu-Jona Lasinio model \citep{Buballa:2003qv}; 2) Color
superconductivity was (re-)discovered. In particular, it has been shown that a phase based on gaps involving color and all three
flavors, called Color-Flavor Locked (CFL) phase, could be the absolute ground state of matter, at least at very large densities \citep{Alford:2007xm}. Some of the calculations were based on MIT-bag-like models, but most were based on extensions of chiral models. In the latter, the large mass of the strange quark makes its production difficult even at the center of the most massive compact stars \citep{Buballa:2003qv,Alford:2007xm}; 3) High order perturbative calculations of dense quark matter were performed, also incorporating strange quarks and stressing that SQM is not soft and it allows for having compact stars with masses largely exceeding 2$\mathrm{M}_\odot$ \citep{Alford:2007xm,Kurkela:2009gj}. 

What is the theoretical status of the EoSs that can support the BW hypothesis? While it has been known for many years
that MIT-bag-like models can satisfy the requests at the basis of that hypothesis \citep{Farhi:1984qu}, it has been shown that
NJL-like models are incompatible with it \citep{Klahn:2015mfa}. At the moment, it has been demonstrated that the BW
hypothesis can also be realized in the color-dielectric model \citep{Drago:2001nq,Alberico:2001zz,Dondi_2017}
and, more generally, in models in which quark masses have a strong density dependence related to confinement. A crucial role in these analyses can be played by the gluon dynamics in the infrared regime \citep{Cyrol_2018,HUBER20201}. In particular, the gluon condensate can act as an effective gluon mass, regularizing the infrared behavior \citep{Adler:1983zh,Horak:2022aqx}. The BW hypothesis can also be realized in models based on perturbative expansions \citep{Kurkela:2009gj}.

In 1984, when Witten wrote his seminal paper, it was considered very plausible that the primordial confinement/deconfinement phase transition was first order. Lattice QCD calculations have later investigated the order of the
phase transition in the regime of small baryonic and leptonic chemical potentials, and it has been found that it corresponds to a smooth crossover \citep{Aoki:2006we}.It is still expected that a critical endpoint exists, in the chemical potential - temperature plane, and that to the right of that point the transition is a first order \cite{Stephanov:2004wx}. Recently, it has been shown that, if large leptonic asymmetries are taken into account (which is not excluded by cosmology \citep{Froustey_2024}), the trajectory followed by the Early Universe can indeed pass through a first order transition \citep{Gao:2021nwz}. It is not yet clear if this will
open again the possibility of forming strangelets in the Early Universe following the scheme outlined by Witten. Finally, an important consequence of a first order phase transition in the Early Universe is the production of gravitational waves \citep{Witten:1984rs} which could contribute to the signal recently detected by the NANOGrav collaboration \citep{Afzal_2023}.

Moreover, at LHC experiments such as ALICE, observations have been made of the formation of light nuclei and hypernuclei during proton-proton and nucleus-nucleus collisions \citep{ALICE:2015oer,Braun-Munzinger:2018hat}, even though the temperature at the freeze-out is significantly larger than the scale of their binding energy. This rather surprising result indicates that the process of clusterization can start at those large temperatures and that this process can exist even in the absence of a first-order phase transition. Actually, the process might be based on percolation \citep{Martens:2006ac}, a transition that is, in general, second order \citep{christensen2005complexity}. It has been found that correlations are present
between hadrons at temperatures close to the critical one \citep{Bellwied:2021nrt}, which might suggest
that clusterization can indeed start, maybe using the formation of the H dibaryon \citep{Jaffe1977}
(or sexaquark \citep{Farrar:2022mih,Shahrbaf:2022upc}) as the doorway towards the formation of larger clusters. Besides, there are indications that the transition temperature might be flavor-dependent, with strange quarks hadronizing at higher temperatures, compared to light ones \citep{Bellwied:2013cta,Alba:2020jir,Flor:2020fdw}. This might lead to the early formation of strange clusters in the Early Universe. In conclusion, even if a first order confinement phase transition does not take place in the Early Universe, it is possible that cosmological strangelets can form following mechanisms similar to the ones responsible for the formation of nuclei in ALICE.

In this paper we will shortly revisit the limits on strangelets coming from a variety of astrophysical, cosmological and laboratory observations. We will then produce a set of mass distributions for strangelets, based on a simple scheme similar to the one discussed by \citet{Witten:1984rs} and including the partial evaporation of strangelets as discussed by \citet{Alcock:1985vc} and revisited by \citet{Madsen:1986jg}. We will fix the few parameters of the model to satisfy all the constraints mentioned above. While this mechanism is not necessarily the correct one, our purpose is to show that it is possible to obtain mass distributions of strangelets that do not violate any observational limit, and based on a simple hadronization scheme.

If a credible distribution of the strangelet masses is obtained, it is possible to estimate their capture rates on the main sequence and on compact stars. Recently, some of us started exploring this idea in connection with possible processes leading to the formation of compact objects having masses of the order of or smaller than one solar mass \citep{DiClemente2022_hess,DiClemente:2022ktz}, as suggested in \citet{DiSalvo:2018mua} and in \citet{Doroshenko2022}, The possible implications of a dark matter made of SQM on the stellar evolution are certainly much broader and will require more extensive analyses. 

Another problem that can be tackled once a candidate distribution of strangelets is at hand, is the formation of supermassive black holes. It has been suggested that self-interacting dark matter can form black hole seeds in the center of a halo, potentially solving the problem \citep{Pollack:2014rja}. We will show that strangelets do not satisfy the requests on self-interaction needed to solve that problem, but on the other hand they interact with ordinary matter and the implications of that interaction still need to be explored.

Finally, if the mass distribution of strangelets is obtained, it is possible to formulate well-focused campaigns to search for these objects. In the last part of this manuscript, we discuss possible search strategies to find the strangelets with the mass distribution that we predict.

\section{Observational constraints}\label{obsconstr}
Dark matter could be either intrinsically weakly interacting, or effectively weakly interacting due to its large mass and, consequently, its low number density \citep{Jacobs:2014yca}. The primary candidates for dark matter have been for a long time Weakly Interacting Massive Particles (WIMPs), but the lack of conclusive evidence from collider experiments and from campaigns of direct detection has left their existence uncertain. 

Strangelets belong to the macros category, which has to respect several observational constraints from different sources \citep{Jacobs:2014yca,Burdin:2014xma}. The mass ranges not subject to limits for nuclear density monochromatic macros in the present Universe are $(55-10^{3})$ g, $(5\times 10^4-10^8)$ g and $(10^{10}-10^{18})$ g \citep{SinghSidhu:2019tbr,Burdin:2014xma}. In the case of strangelets \footnote{Other types of macros composed of quarks/antiquarks have been proposed, for instance the ones associated with the collapse of domain walls \citet{Bai:2023cqj} and \citet{Liang:2016tqc}.}, other constraints must also be taken into account, because they can interact with ordinary matter, dramatically altering the astrophysical evolution of stars. In particular, \citet{Madsen:1988zgf} rules out the two lowest mass windows, leaving open (at least for monochromatic strangelets) only the range $(10^{10}-10^{18})$ g.

Here follows a short list of the main current observational constraints.
\subsection{Ancient Mica}
Strangelets with low enough mass, would have a high enough number density to have left a record on Earth. If the mass was not very large, they could have penetrated a few kilometers into the Earth's crust, thus leaving an imprint into ancient Muscovite mica.
In examining the potential impact of macros on ancient mica, the study of \citet{Jacobs:2014yca} accounts for the absence of traces found in the sample considered by \citet{Salamon:1986} and it excludes strangelets having a mass $M<55$ g.

\subsection{Seismic events}
Limits on the mass have been set by considering seismic events on Earth and on the Moon \citep{Burdin:2014xma,Herrin:2005kb}. Concerning seismic events on Earth, the mass range  $10^{5}\,\mathrm g <M < 3\times10^{8}\,\mathrm g$  as dominant dark matter component can be excluded. Moreover, from the measurements of the total amount of seismic energy, estimated by using the seismic station on the Moon implanted during the Apollo missions, it is possible to conclude that less than 10$\%$ of dark matter density is in the mass range $5\times10^{4} \,\mathrm g<M < 10^{6} \,\mathrm g$. 

\subsection{Femtolensing}
One of the ways to search for compact astrophysical objects that could constitute dark matter, is through the lensing effects that they would cause. Femtolensing is an interference effect in the energy spectrum of the lensed object. Using Fermi satellite data, \citet{Barnacka_2012} have set a limit of contribution to dark matter mass of 8$\%$, for compact objects having masses in the range of $5 \times 10^{17}\,\mathrm g <M< 2 \times10^{20}\, \mathrm g$. Nevertheless, the more recent analysis from \citet{Katz:2018zrn} which revised femtolensing, states that there is no such a limit.

\subsection{Microlensing}
The analysis of \citet{SinghSidhu:2019tbr} rules out macros in the range of mass
$5\times10^{22} \,\mathrm g<M < 4\times 10^{24}\,\mathrm g$ 
as the dominant component of dark matter. Moreover, \citet{Jacobs:2014yca} rules out macros having $M\geq 10^{24}\text{ g}$.

\subsection{Thermonuclear runaway}
A macro of sufficiently large cross-section impacting a white dwarf (WD) or a neutron star would release an energy leading to a thermonuclear runaway, therefore a supernova explosion. Based on this assumption, the \citet{SinghSidhu:2019tbr} analysis excludes the mass ranges  $10^{8} \mathrm g<M < 10^{10} \mathrm g$  and $M>10^{18}$ g for nuclear density macros.

\subsection{Proto-neutron stars}
Not all compact stars are QSs \citep{Watts:2006hk}. A strangelet can penetrate a newly formed NS, before a crust forms, and convert it into a QS. Therefore, one needs to impose a limit on the strangelet flux. Given $\tau_\mathrm{melt}$ as the time before the NS forms a crust able to stop strangelets \citep{Madsen:1988zgf} and given that only strangelets with a baryon number $A > 10^{12}$ can evade being trapped by the expanding supernova shell, the relation $F_{12} \times\tau_\mathrm{melt} \ll 1$ must hold, where $F_{12}$ is the flux of strangelets having $A > 10^{12}$.

\subsection{Comparison with primordial black-holes}\label{subsec:comparisonPBH}
Primordial black holes (PBHs) have long been considered as potential dark matter candidates \citep{Carr:1974nx,Carr:1975qj,Carr:2020gox}, and many of the observational limits that restrict strange quark matter as a dark matter component also apply to PBHs. Constraints from gravitational microlensing surveys such as MACHO, EROS and OGLR \citep{Alcock:2000ph,EROS-2:2006ryy,Niikura:2017zjd}, and from femtolensing \citep{Barnacka_2012} shape the allowed mass ranges of both PBHs and strange quark nuggets.

Despite these parallels, one crucial difference is due to the possible evaporation of PBHs. Black holes with masses smaller than $\sim 10^{17}\,\mathrm{g}$ are effectively ruled out as significant dark matter constituents due to Hawking radiation \citep{Hawking:1974rv,Hawking:1975vcx}. Such low-mass PBHs would have evaporated well before the present epoch, depleting their abundance. In contrast strangelets, do not undergo a Hawking-like evaporation process and therefore they are not subject to this constraint.

\section{A simple phenomenological model}
In the following, we discuss a model for the production of strangelets in the Early Universe, based on the scheme developed in \citet{Witten:1984rs}, and incorporating the evaporation mechanism discussed in \citet{Alcock:1985vc} and revisited by \citet{Madsen:1986jg}. This simple scheme does not discuss many relevant problems which emerged during the forty years after the publication of the paper of \citet{Witten:1984rs}. In particular, the mechanism of chiral symmetry breaking was not incorporated in that scheme, which was loosely modeled on the MIT bag model. In \citet{Farhi:1984qu}, they discussed values of the bag constant $B$, of the strong coupling $\alpha_s$ and of the strange quark mass $m_s$ for which SQM is absolutely stable, but most of that parameter space has now been ruled out, in particular by analyses of heavy ion collisions. On the other hand, the study of the formation of gapped phases has shown that absolutely stable SQM can be obtained in a CFL phase \citep{Alford:2007xm}.

While in the following we discuss a simple scheme, still based on the papers of \citet{Witten:1984rs} and \citet{Alcock:1985vc}, at the end of this section we will propose a possible path for the Early Universe which is compatible with recent microphysics and still leads to the formation of cosmological strangelets.

\subsection{Evaporation}\label{sec:evaporation}
\citet{Witten:1984rs} discusses the typical size of cosmological strangelets, formed at the time of hadronization, suggesting that it could be of the order of centimeters. As discussed above, that analysis needs to be revisited, and we will simply assume that the initial size of the strangelets is regulated by a distribution, that we choose to be log-normal, as suggested by studies on the coalescence of droplets \citep{Davis1973}. A log-normal distribution has also been used to describe the sizes of coalescing primordial black holes \citep{Dolgov_2020}. This distribution will be modified by the process of evaporation. 

The evaporation of SQM, as discussed by \citet{Alcock:1985vc}, is a crucial phenomenon to understand the behavior of matter in the Early Universe. In particular, when the Universe cools from $T\sim150$ MeV to $T\sim1$ MeV, the evaporation of SQM plays an important role in determining the survival and stability of strangelets. 

The stability of strangelets is highly dependent on their baryon number. \citet{Alcock:1985vc} suggest that strangelets, with a baryon number smaller than a critical one, evaporate completely in the Early Universe. Notice anyway that strangelets having a baryon number larger than the critical one undergo only a partial evaporation, and the final distribution of their sizes is therefore the result of this process. 

The mechanism of evaporation primarily involves the emission of nucleons from the strangelet.
This process is regulated by various mechanisms, as outlined in \citet{Alcock:1985vc}, in particular: the thermal balance of the strangelet which cools down by the evaporation process itself and needs to be kept warm by neutrinos; the net-rate of emission of nucleons by an isolated strangelet; the increase in the re-absorption rate due to the presence of evaporated nucleons close to the surface of the strangelet; the opacity of the evaporated material to neutrinos. More in details:

\begin{itemize}
    \item the primary consideration is the energy balance, which includes the cooling effects due to evaporation and neutrino emission, and the reheating effects from incoming neutrinos. The probability of neutrino absorption is contingent on the neutrino's mean free path $l(T)$, the strangelet size $r_s$ ($\propto A^{1/3}$), and the environment temperature $T$. 
 If the size of the lump exceeds the mean free path of a neutrino, it becomes opaque to neutrinos and will emit them with a thermal spectrum. The energy emission rate of strangelets, determined by its surface temperature $T_s$, is given by:
\begin{equation}
L_{\mathrm{th}}=4 \pi r_s^2 (7 \pi^2 /160)T_s ^4
\end{equation}
indicating that the lump radiates according to its temperature $T_s$. On the other hand the absorption rate reads:
\begin{equation}
L_{\mathrm{abs}}=4 \pi r_s^2 (7 \pi^2 /160)T_u ^4,
\end{equation}
where $T_u$ is the temperature of the Universe. For a strangelet smaller than the neutrino mean free path, the emission and the absorption rates should be adjusted by incorporating a probability factor:
    \begin{equation}
p\left(r_{s}, T\right)= \begin{cases}1, & r_{s}>\frac{3}{4} l(T) \\ \frac{4 r_{s}}{3 l(T)}, & r_{s} \leq \frac{3}{4} l(T).\end{cases}
    \end{equation}
Therefore, the total rate reads:
\begin{equation}
L=4 \pi r_{s}^{2}\left(\frac{7 \pi^{2}}{160}\right)\left[T_{u}^{4}\, p\left(r_{s}, T_{u}\right)-T_{s}^{4}\, p\left(r_{s}, T_{s}\right)\right]\label{luminositystrg} \, .
\end{equation}

\item By solving a detailed balance equation, it is possible to estimate the net emission rate of nucleons from an isolated strangelet:
\begin{equation}\label{neutronabs}
    r_{n,p}=\frac{m_{N} T_{s}^{2}}{2 \pi^{2}}\, e^{-I / T_{s}} \,f_{n,p}\, \sigma_{0} A^{2 / 3},
\end{equation}
where $f_n$ ($f_p$) are the neutron (proton) absorption efficiencies, $I$ is the binding energy, $\sigma_0\, A^{2 / 3}$ is the geometrical cross section and $\sigma_0=3.1\times10^{-4}$ MeV$^{-2}$ is obtained by assuming a density of (125 MeV)$^3$, while $m_{N}$ is the nucleon mass. 
   
\item 
It is necessary to consider that nucleons close to the surface of strangelets might be reabsorbed \footnote{At the high temperatures relevant for  evaporation the Coulomb barrier surrounding the strangelets does not stop the protons from being absorbed, see discussion in \autoref{rephrasingwitten}}. The rates at which neutrons and protons are reabsorbed are as follows:
\begin{equation}
    r_{n,p}^{\text{abs}}=f_{n,p}\, N_{n,p} \,\sigma_{0} \,A^{2/3} (T_s/2 \pi m_N)^{1/2}
\end{equation}
where $N_n$ ($N_p$) are the neutron (proton) densities, which depend on the temperature.

    \item 
    Local densities of neutrons and protons are determined by pressure equilibrium. We take into account the contribution of electrons, positrons, photons, nucleons, and the contribution of neutrinos since we assume that they interact with the evaporated material:
\begin{equation}
    \left(N_{n}+N_{p}\right) T+\frac{43 \pi^{2}}{360} T^{4}=\frac{43 \pi^{2}}{360} T_{u}^{4}.
    \end{equation}
Including all emission and absorption terms, the net-evaporation rate of a cosmological strangelet can be formulated as:
\begin{equation}
L=\frac{d A}{d t}\left(I+2 T_{s}\right)\label{evapoeq1} \, ,
\end{equation}
where:
\begin{equation}
\begin{aligned}
\frac{d A}{d t}= & \left[\frac{m_{N}\, T_{s}^{2}}{2 \pi^{2}} e^{-I / T_{s}}-\frac{43 \pi^{2}}{720 T_{s}}\left(T_{u}^{4}-T_{s}^{4}\right)\left[\frac{T_{s}}{2 \pi m_{N}}\right]^{1 / 2}\right]\\
& \times \sigma_{0}\, A^{2 / 3}\left(f_{n}+f_{p}\right).
\end{aligned}\label{evapoeq2}
\end{equation}
Note that $dA/dt$ is the evaporation rate, a positive quantity, and it has the opposite sign with respect to the rate of change of the baryon content of the strangelet. 
By combining \autoref{luminositystrg}, \autoref{evapoeq1}, and \autoref{evapoeq2} one can obtain an equation relating $T_u$, $T_s$ and $A$:
\begin{align}
    &\frac{k_1 \, \left[T_u^4 p\left(T_u,A\left(T_u\right)\right)-T_s^4 p\left(T_s,A\left(T_u\right)\right)\right]}{\left(2 T_s+I\right) \left(f_n+f_p\right)}\nonumber \\&= -\frac{k_2\left(T_u^4-T_s^4\right)}{\sqrt{T_s}}+\frac{m_N T_s^2 e^{-\frac{I}{T_s}}}{2 \pi ^2}.\label{evapobeta2}
\end{align}
\item The evaporated material surrounding the strangelets can be thick to neutrinos. At high temperatures, neutrinos are absorbed by that external layer and cannot reach the strangelets. In that situation, evaporation is suppressed. This limiting temperature $T_\mathrm{evap}$ is indicated in Fig. 1 of \citet{Alcock:1985vc}, it depends on the mass of the strangelet and it is in the range $50 \,\mathrm{MeV}\lesssim T_\mathrm{evap} \lesssim 120\,\mathrm{MeV}$. We will consider $T_\mathrm{evap}$ as a free parameter.

Finally, given the relation between the age of the universe $\tau_U$ and its temperature $T_u$ in the radiation-dominated era \citep{Alcock:1985vc}:
\begin{equation} 
    \tau_U = \left(\frac{45}{172\pi^3 G}\right)^{1/2} \frac{1}{T_u^2}\,,
\end{equation}
it is possible to write the evaporation rate as a function of the temperature $T_u$ alone:
\begin{equation}
    \frac{dA}{dT_u} = -
    \frac{ \left(\beta\,  k_3\, A^{2/3} \left[T_u^4 \,p\left(T_u,A\right)-T_s^4 \,p\left(T_s,A\right)\right]\right)}{\left(2 T_s+I\right)T_u^3}. \label{evapobeta}
\end{equation}
Here and in \autoref{evapobeta2}, $k_1$, $k_2$ and $k_3$ are parameters that incorporate various physical constants into the model. Additionally, we introduce a phenomenological factor $\beta$ to account for corrections related to Pauli blocking, to the finite value of the strange quark mass and to QCD effects. These corrections can suppress the evaporation rate by several orders of magnitude \citep{Madsen:1986jg,Madsen:1988zgf}. By solving \autoref{evapobeta2} and \autoref{evapobeta}, one can effectively determine the dynamics of strangelet evaporation during the cooling phase that follows hadronization.

\end{itemize}
\subsection{Strangelet distributions}
To accurately model the current distribution of dark matter in the galaxy as strangelets, one must consider various factors, including the strangelet number distribution. Initially, we hypothesize a log-normal distribution of sizes and then subject it to an evaporation process. This evaporation, occurring in a range of temperatures from 150 MeV to 1 MeV, results in a transformed distribution that must agree with the constraints specified in \autoref{obsconstr}. 

\subsubsection{Mathematical framework}

\begin{figure}
\begin{centering}
\includegraphics[width=\columnwidth]{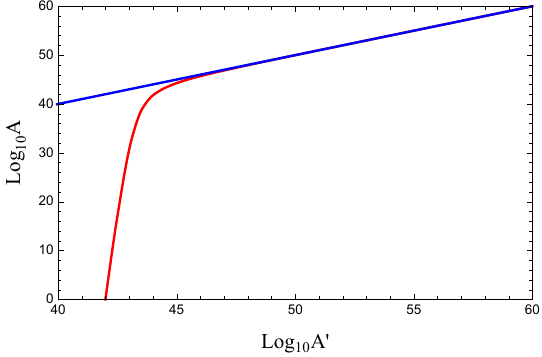}
    \caption{Red curve: example of the function $A=f(A')$, which maps $A'$, the size of the strangelets pre-evaporation, onto the post-evaporation size $A$. In this example, strangelets having $A' \lesssim 10^{42}$ evaporate completely. It is compared with the blue curve, which corresponds to the case in which there is no evaporation.}
    \label{evapor}
\end{centering}
\end{figure}

It is crucial to establish a mathematical framework for handling the number distributions of strangelets. Here we will follow the scheme developed in \citet{Bucciantini:2019ivq}. Given an initial distribution $P(A')$, representing the number of strangelets with a baryon number $A'$, the distribution after evaporation, denoted as $Q(A)$, is given by:

\begin{equation}
    Q(A)=\int_1^\infty \mathrm{d}A' P(A') \delta (A-f(A')).
\end{equation}
Here, $f(A')$ is the function that maps the pre-evaporation baryon number $A'$ onto the post-evaporation baryon number $A$, see \autoref{evapor}, and it can be evaluated by using the technique outlined in the previous section. 

It is important to remark that the number of strangelets is not conserved during the process of evaporation, since those having a size smaller than a critical one evaporate completely:
\begin{equation}
\int_1^\infty Q(A) \, \mathrm 
d A < \int_1^\infty  P(A) \, \mathrm d A  
\end{equation}
This framework allows us to map the initial distribution of the baryon number of strangelets onto the distribution after evaporation. In this way, the total amount of baryons, evaporated from the strangelets, will provide the baryon number of ordinary matter, allowing us to achieve consistency with observational constraints.

\subsubsection{Parameters of the model}
\begin{figure}
\begin{centering}
\includegraphics[width=\columnwidth]{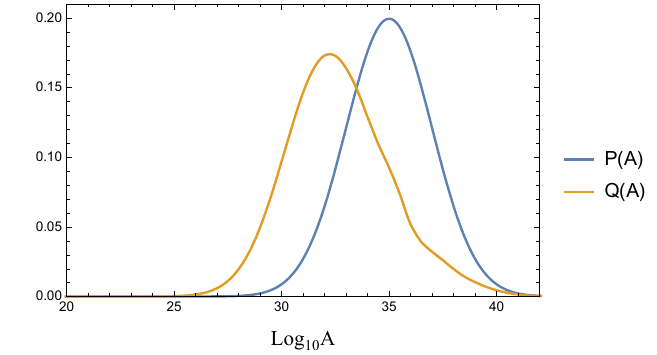}
    \caption{Example of the relation between the pre-evaporation distribution $P(A)$ and the final distribution $Q(A)$. $P(A)$ is a log-normal distribution with $\mu_{\text{max}}=35$, $\sigma=2$. $Q(A)$ is obtained by evaporation, by assuming $\text{log}_{10}\,\beta=-4.7$ (see \autoref{evapobeta}). Here $P(A)$ is normalized to 1.}
    \label{evaporized}
\end{centering}
\end{figure}

As discussed above, we assume that the initial number distribution of strangelets having a given mass, at the time of hadronization and before evaporation, is log-normal. 
It reads:
\begin{equation}
f(A; \mu, \sigma) = \frac{1}{A\sigma\sqrt{2\pi}} \exp\left( -\frac{(\ln A - \mu)^2}{2\sigma^2} \right), 
\end{equation}
where $\mu$ and $\sigma$ are the mean and the standard deviation of the variable's logarithm, not of the variable $A$ itself. Notice that the maximum of the log-normal is located at $\ln A = \mu -\sigma^2\equiv \mu_\mathrm {max}$.
In \autoref{evaporized} we show an example of the initial distribution $P(A)$ and of the post-evaporation distribution $Q(A)$.

In our scheme we have therefore four parameters regulating the final distribution $Q(A)$: the mean $\mu$ and the standard deviation $\sigma$ of the log-normal distribution; the temperature $T_\mathrm{evap}$ at which neutrinos can reheat the strangelets and evaporation becomes efficient; the parameter $\beta$, regulating the evaporation rate. 

To evaluate the viability of this model, several constraints must be considered. These include ensuring that the final distribution aligns with the observed ratio of dark matter in the Milky Way, which is approximately 90\% of the total matter content \citep{Watkins_2019}. Additionally, the model must adhere to the observational constraints outlined in \autoref{obsconstr}. Through a systematic exploration of the parameter space, we correlate the initial distribution parameters with those of the final distribution. This approach enables us to control and predict key characteristics of the strangelet population, including the maximum of the final distribution of the baryon number $Q(A)$ and of the mass $A\,Q(A)$, and the final number of strangelets $N\equiv\int Q(A) \mathrm d A$.

In Appendix \ref{paramspace} we analyze the impact of these four parameters on the final distribution of strangelets. An important outcome of that analysis is that, if we reduce the value of $T_\mathrm{evap}$ i.e. if we assume that evaporation starts being effective at lower temperatures, we get final distributions similar to the ones obtained by reducing the value of $\beta$, i.e. assuming a strong suppression of the decay rate. In particular, we get similar results if we assume $T_\mathrm{evap}\sim 150 \mathrm{MeV}$, $\beta\sim 10^{-5.6}$ or if instead we use $T_\mathrm{evap}\sim 50 \mathrm{MeV}$, $\beta\sim 10^{-4}$. This second choice is probably more realistic in light of the analyses of \citet{Alcock:1985vc} and of \citet{Madsen:1986jg,Madsen:1988zgf}.

\subsubsection{Galactic distribution}
\begin{figure}
\begin{centering}
\includegraphics[width=\columnwidth]{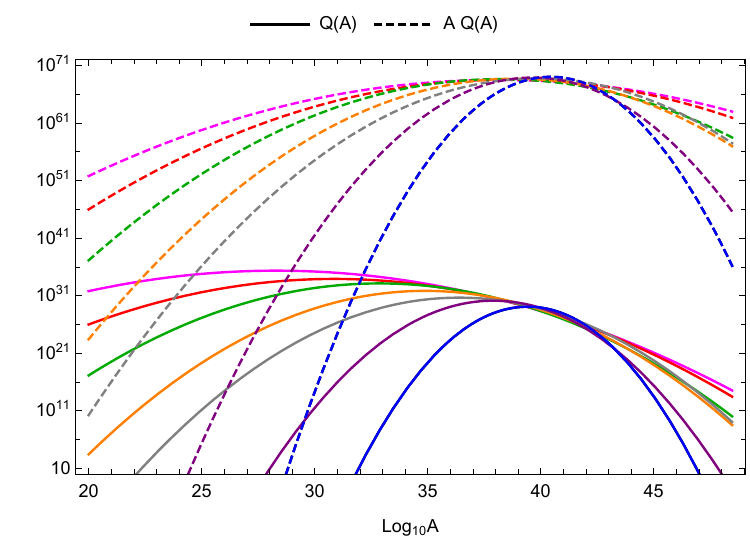}
    \caption{Number distributions (solid) and mass distributions (dashed) of strangelets. The curves in red, green, yellow, gray, purple, and blue, correspond to distributions containing approximately $10^{34}$, $10^{33}$, $10^{32}$, $10^{31}$, $10^{30}$, and $10^{29}$ strangelets, respectively. The distributions are consistent with the observational constraints detailed in \autoref{obsconstr} and have been normalized to the estimated dark matter mass in the Milky Way \citep{Nesti_2013}.}
    \label{distr}
\end{centering}
\end{figure}

By implementing the evaporation model on the initial distributions $P(x)$, and ensuring the agreement with the constraints, we have been able to calculate the number distribution of strangelets and the mass distribution. These distributions are plotted in \autoref{distr}.

Understanding the astrophysical implications of dark matter, particularly if a portion of it is composed of strangelets, necessitates an estimation of the flux of strangelets impacting on celestial bodies. For this purpose, we utilize the NFW profile \citep{Navarro:1996gj}, which is a well-established model in astrophysics, widely used for describing the density distribution of dark matter in galaxies. As described in \citet{Nesti_2013}, the NFW profile is given by
\begin{equation}
\rho_{N F W}=\rho_H \frac{1}{x(1+x)^2},
\end{equation}
where $\rho_H$ represents the characteristic density scale of the profile. The dimensionless variable $x$ is defined as $x=r/R_H$, with $R_H$ being a scale radius. The profile's slope at $R_H$ is characterized by:
\begin{equation}
\left[\frac{d \log \rho_{N F W} }{ d \log r} \right]_{r=R_H}=-2\,.
\end{equation}
\begin{figure}
\begin{centering}
\includegraphics[width=\columnwidth]{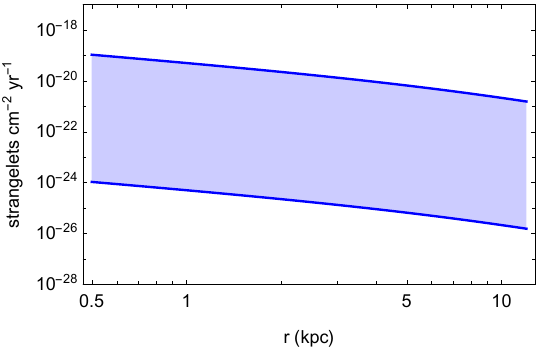}
    \caption{The plot illustrates the expected flux of strangelets, with the number of strangelets ranging from $10^{29}$ to $10^{34}$ (see \autoref{distr}), as a function of the distance from the Galactic center.}
    \label{nfwprofile}
\end{centering}
\end{figure}
Utilizing the parameters estimated by \citet{Nesti_2013}, we can calculate the expected flux of strangelets as a function of the distance from the center of the Milky Way. This calculation is crucial in understanding the spatial distribution and potential impact of strangelets. The results are illustrated in \autoref{nfwprofile}.
Despite its simplicity, this analysis is key in evaluating the feasibility of strangelets as dark matter candidates and their potential observable signatures.

\subsection{Rephrasing Witten with a modern language}\label{rephrasingwitten}
As already discussed, some of the hypotheses assumed in \citet{Witten:1984rs} and \citet{Farhi:1984qu} are probably not compatible with what is now known about microphysics. In particular the following questions were not discussed, mostly because forty years ago these problems were still in a very primordial stage or completely unknown: chiral dynamics across the critical temperature; compatibility of the value of the bag pressure with the results of experiments of heavy ion collisions; role played by the formation of diquark condensates. In this section we discuss a possible path which, although speculative, incorporates state-of-the-art knowledge of quark dynamics  and could lead to the formation of strangelets in a way which is phenomenologically not too different from the one based on the papers of \citet{Witten:1984rs} and \citet{Farhi:1984qu}. It is indeed interesting to remark that the distribution of the sizes of the strangelets, obtained in the previous section by taking into account all observational limits, peaks at sizes very close to the one estimated in the paper by \citet{Witten:1984rs}. It is therefore important to investigate whether a cosmological path, incorporating up-to-date results of microphysics, can still support that result.

\begin{figure}
\begin{centering}
\includegraphics[width=\columnwidth]{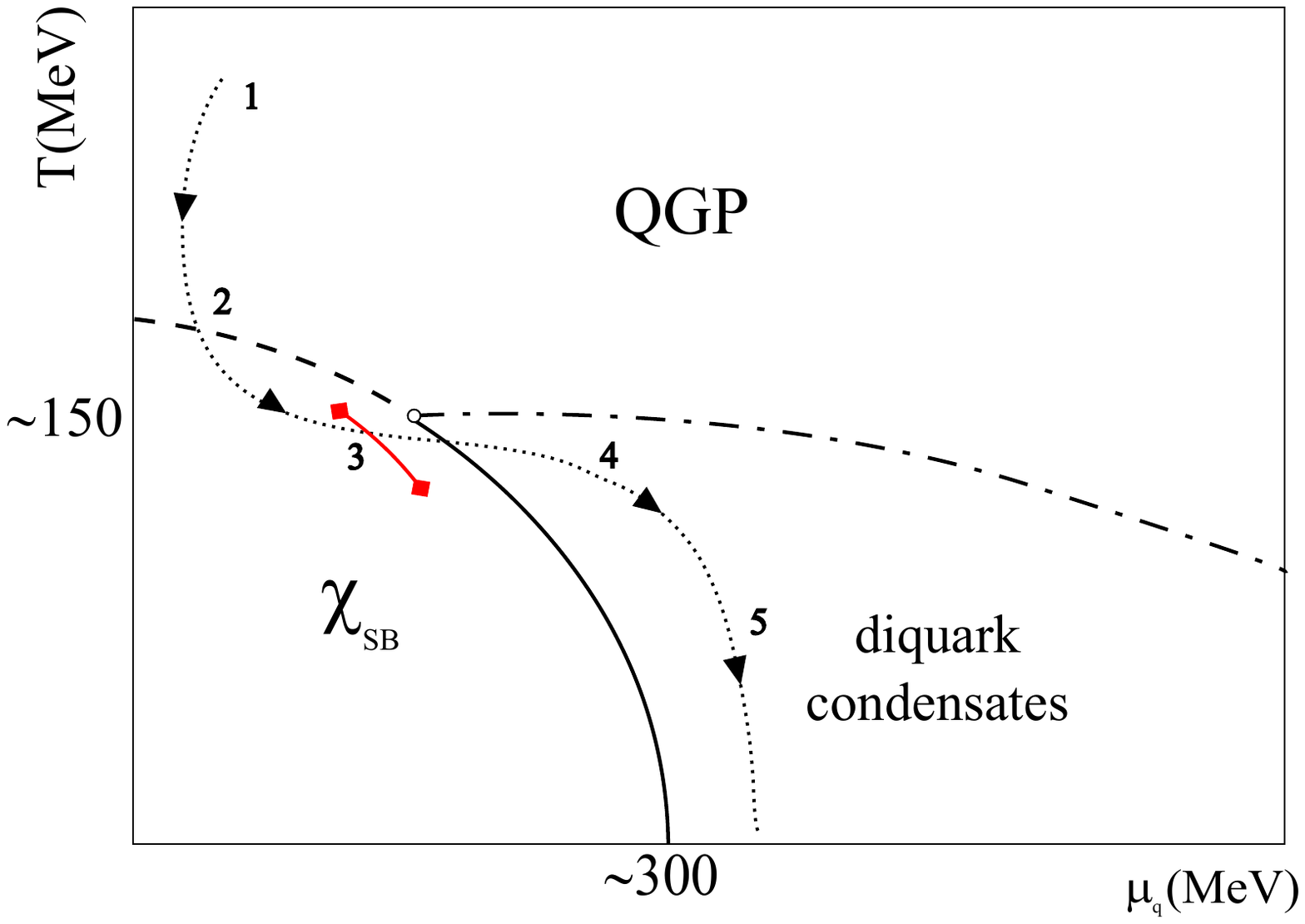}
    \caption{Cosmological path of quark clusters, following the stages outlined in \autoref{rephrasingwitten}. The phase separation is schematic and is loosely based on \citet{Roessner:2006xn}. The solid line indicates a first-order phase transition between the chiral symmetry broken phase and the diquark phase; the dashed line a crossover and the dot-dashed a second order. This last transition, separating the diquark phase from the quark gluon plasma (QGP), can be much more complicated then the schematic one shown in the figure, and its order is strongly model dependent. The stages include: (1) initial quark correlations above the critical temperature, (2) growth of binding energies and quark-coalescence, (3) formation of large clusters, (4) stabilization of the clusters by color superconducting gaps and (5) partial evaporation. The red segment indicates the instability discussed in \citet{Lavagno:2022orw}. If that instability exists and produces hadronic matter with a finite and negative strange charge it is possible that the transition between hadronic matter and CFL phase is totally continuous, see text.}
    \label{phasediagram}
\end{centering}
\end{figure}

We propose a scheme based on the following stages, see \autoref{phasediagram}:
\begin{enumerate}
    \item lattice QCD simulations show that, at temperatures larger than the critical one, strong correlations between quarks can already take place \citep{Bellwied:2021nrt}. Notice that, at this stage, chiral symmetry is still restored and therefore the masses of up, down and strange quarks are still small;
    \item while the transition is not assumed to be first order, we imagine that the decrease of the temperature can lead to an increase of the correlations and to the gradual development of binding energy in the cluster, with a mechanism similar to quark-coalescence \citep{song2022,LHCb:2023wbo}. Once the cluster is bound, its total energy can be described by a semi-empirical mass formula, including also a surface term. If the cosmological trajectory is not too far from the critical end point \citep{Gao:2021nwz}, these correlations can develop more easily;
    \item the size of the clusters, initially of the order of few femtometers, can dramatically increase, by a process of cluster coalescence totally similar to the one discussed by \citet{Witten:1984rs}. This process could start with the fusion of clusters of three quarks, maybe by using clusters having the quantum numbers of the sexaquark as the doorway to the formation of much larger clusters \citep{Farrar:2022mih,Shahrbaf:2022upc}. This process should be mildly exothermic and therefore the temperature should remain roughly constant. A potential problem with this path is that, unless the new phase is stable under weak interactions, it can decay on a timescale shorter than the one needed to form large clusters.  
    Another possible path leading to the formation of larger clusters is based on the existence of chemical-mechanical instabilities, in the hadronic phase, at temperatures close to the critical one. This possibility has been discussed in \citet{Lavagno:2022orw}, where it has been shown that a phase separation can take place, between a low-density, positive-strange-charge phase and a higher density phase with a negative strange charge. This second phase can form, on the timescale of the strong interaction, as soon as the system starts hadronizing, it has a finite content of hyperons and it can constitute the bridge to the formation of diquark matter described in the next stage. The exact location of this instability still needs to be worked out (and is model dependent), but it could start at a chemical potential smaller than that of the QCD critical point (see red segment in \autoref{phasediagram}). This is not surprising, since it is known, also from lattice calculations, that phase transitions in the hadronic phase can take place at very low densities, for instance for isospin asymmetric matter \citep{Brandt:2017oyy,Vovchenko:2020crk}; 
    \item even if in the previous stage large clusters can already form, they can be meta-stable and therefore rapidly evaporate. This problem could be solved by the formation of color superconducting energy gaps \citep{Alford:2007xm,Lugones:2003un}, but the maximum temperature at which these gaps can form is often estimated to be significantly smaller than the deconfinement temperature. The existence of a rather large gap between these two temperatures would lead to a rapid evaporation of the clusters. This problem was noticed by \citet{Zhitnitsky:2002qa} and led to the suggestion that the clusters could be stabilized by the pressure of axion domain walls.
    {\it{We propose instead that the value of the superconducting gap, and therefore of the associated critical temperature, is significantly larger than the one normally discussed in the literature.}} Notice that most of the estimates present in the literature are based on the NJL model \citep{Ruester:2005ib}, but a similar calculation based on the PNJL model suggests that the diquarks can form at temperatures not much smaller than the critical one \citep{Roessner:2006xn}. Also a preliminary calculation within the chromodielectric model suggests significantly larger gaps, whose value increases at low densities \citep{deCarvalho:2010wdj}. At variance with NJL, PNJL and the chromodielectric are confining models and this can play a crucial role in order to obtain absolutely stable strange quark matter \citep{Drago:2001nq,Dondi_2017}.  
    Notice that since up, down and strange quarks are already present in the cluster, nothing stops the cluster from forming a phase in which quarks are in a CFL phase. This process can take place only if the clusters are larger than several femtometers, otherwise the gap formation is suppressed \citep{Amore:2001uf}. At that point, the clusters are totally stable and they follow the partial evaporation path described in the previous section. This mechanism of stabilizing clusters of SQM has been addressed in several papers, including the ones discussing the evolution of an entire strange quark star \citep{Alford:2002rj,Drago:2004vu}. Stages iii and iv can be summarized by saying that we assume that a phase transition takes place, associated with a mechanical instability, which leads to the formation of a new phase having a large baryon density and a finite and negative strange charge. This new phase can rapidly transform into a CFL phase, which can be particularly easy if there is no phase transition between hypernuclear matter and CFL quark matter \citep{Sch_fer_1999,Alford_1999,Alford:2007xm}. In this way we have re-introduced a phase transition with a large density jump, similar to the one discussed by \citet{Witten:1984rs}, but taking place at temperatures slightly smaller than the critical one in the already hadronized matter. Also in the scenario discussed here, gravitational waves could be generated during this process \citep{Witten:1984rs,EPTA_2024}.
     
    \item 
    
     The process of evaporation is initially suppressed because neutrinos (crucial to evaporation, as discussed in the previous section) are absorbed by the layer of evaporated hadrons and they are therefore unable to keep the cluster hot \citep{Alcock:1985vc}. On the other hand, evaporation is crucial to fix the right balance between ordinary matter (the evaporated part) and dark matter (the remaining large clusters). 
    
\end{enumerate}

The scheme outlined above is compatible with the evaporation process described in the previous section. In particular, the formation of diquark condensates at large temperatures is crucial to avoid the rapid evaporation of the clusters.

\begin{figure}
\begin{centering}
\includegraphics[width=\columnwidth]{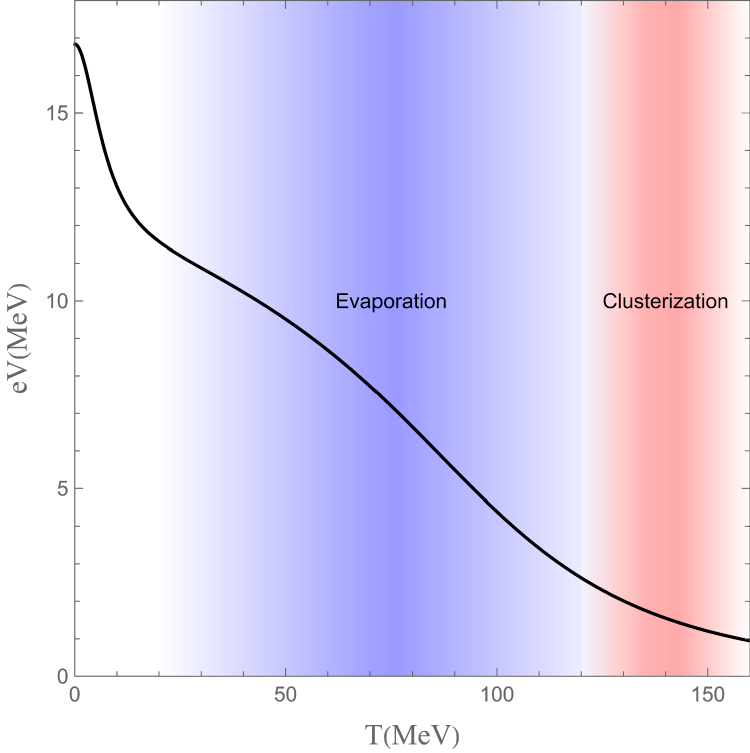}
    \caption{Coulomb barrier for a particle of unitary charge $e$ as a function of the temperature $T$.
    The calculation has been performed by using the MIT bag model, setting the bag constant to $B=145^4$ MeV$^4$ and with quark masses: $m_\mathrm{up}=1$ MeV, $m_\mathrm{down}=1$ MeV, $m_\mathrm{strange}=150$ MeV.}
    \label{coulombbarrier}
\end{centering}
\end{figure}

Up to now we have not discussed the role possibly played by the Coulomb interaction during coalescence. A macroscopic isolated strangelet develops a Coulomb barrier, similarly to what happens in the case of a strange quark star \citep{Alcock:1986hz}. The height of the barrier depends on the temperature \citep{Kettner:1994zs} and in \autoref{coulombbarrier} we show this dependence. At low temperatures ($T\lesssim 1$ MeV) the barrier is large enough to stop the absorption of protons. Instead, for $T\sim$ a few ten MeV (the range relevant for evaporation), the barrier is already low compared to the temperature and therefore it can be neglected when discussing proton re-absorption (see \autoref{sec:evaporation}). Finally, for $T\gtrsim 100$ MeV at which clusterization still takes place, the Coulomb barrier has not yet formed and the value indicated in \autoref{coulombbarrier} corresponds just to an upper limit.

Of course the scheme discussed above is speculative even though the formation of nuclei in heavy ion collisions \citep{ALICE:2015oer,Braun-Munzinger:2018hat} somehow supports this scheme, since it shows that the formation of bound clusters can take place even if a first order phase transition is not observed. Detailed analyses will be needed in order to clarify all the steps. For instance, it is crucial to understand the evolution of the clusters with the temperature. Also, the cosmic trajectory has, up to now, been computed assuming a small contribution of multistrange particles \citep{Letessier:2002ony}. It will be interesting to investigate if some of the models used to describe the formation of nuclei at high temperature, as observed in heavy ion collision experiments, can be adapted to the cosmological scenario. Another important point has to do with the formation of gapped quark matter. We do not know if the gapping will directly produce Color-Flavor-Locked quark matter \citep{Alford:2007xm}, or if it will form, at least as an intermediate step, a crystalline phase as the one described in \citet{Anglani:2013gfu}. Notice that the knowledge of which gapping pattern will actually take place is important also for the phenomenology of dark matter, since Color-Flavor-Locked quark matter is totally neutral, while a crystalline phase needs the presence of electrons to achieve charge neutrality.

\section{Astrophysical implications}
We can study the impact of the possible existence of strangelets by estimating the capture rate onto astrophysical objects as main sequence stars. \citet{Madsen:1988zgf} has already discussed the conditions to have interaction or capture of a strangelet by a star. It is expected that a strangelet would pass through a star like the sun without being captured, given its velocity and cross section.

In order to estimate the capture rate from a main sequence star, we use the flux formula from \citet{Jacobs:2014yca}, opportunely modified to take into account the strangelet number and distribution:

\begin{align}
F=&2.7 \times 10^{29} \,\frac {\rho_\mathrm{NFW}(r)}{\rho_\mathrm{DM}}\, \frac {N} {M_\mathrm{V}} \frac{M}{\mathrm{M}_\odot}\, v_{250}\, \left(\frac{R_c}{\mathrm{R}_\odot}\right)^2 \nonumber \\ &\times \left(1+ 6.2 \left(\frac{\mathrm{R}_\odot}{R_t}\right) \,  v_{250}^{-2}\, \left(\frac{M_t}{\mathrm{M}_\odot}\right) \right),\label{eq:jacobs}
\end{align}
where $N$ is the total number of strangelets in the galaxy, $M_\mathrm{V}$ is the virial mass of the Milky Way in units of the proton mass, $v_{250}$ is the velocity of DM in units of 250 km/s, $M$ is the mass of the star, R$_c$ is the radius of the star core (which is the part of the star capable of stopping strangelets \citep{Madsen:1988zgf}). Indeed,  we have extended the formula originally proposed by \citet{Jacobs:2014yca} to account for the fact that only strangelets colliding with the dense central region of a star, characterized by a radius $R_c$, can be halted, while others may pass through.

To estimate the capture on a star, we have to consider a star which has already formed a core dense enough to stop a strangelet. Indeed, we choose the silicon formation because the mass density would be enough to start stopping some strangelets. Therefore, in order to estimate the capture probability,  \autoref{eq:jacobs} has to be multiplied by $\tau(M)$, which is the time interval between the silicon formation in the core of a star having mass $M$ and its collapse.

\subsection{Strange Dwarfs}
An astrophysical implication of the presence of cosmological strangelets is the impact they can have on the evolutionary path of WDs which, capturing strangelets in their lifetime, become strange dwarfs (SDs). The SD existence and stability have been argued several times \citep{Glendenning:1994sp,Glendenning:1994zb,Vartanyan:2009zza,Vartanyan:2012zz,Alford:2017vca,Kurban:2020xtb,DiClemente:2022ktz}. 

\citet{Glendenning:1994sp} analyzed the radial stability of SDs, suggesting they could be stable even for density of nuclear matter exceeding the maximum for regular WDs. However, \citet{Alford:2017vca} challenged this view. They found that the fundamental radial mode eigenvalue was negative, implying instability in such systems. This apparent contradiction is solved in \citet{DiClemente:2022ktz}, addressing the study of the boundary conditions \citep{Pereira:2017rmp,DiClemente:2020szl} at the interface between nuclear matter and the quark core.

\begin{figure}
\begin{centering}
\includegraphics[width=1.05\columnwidth]{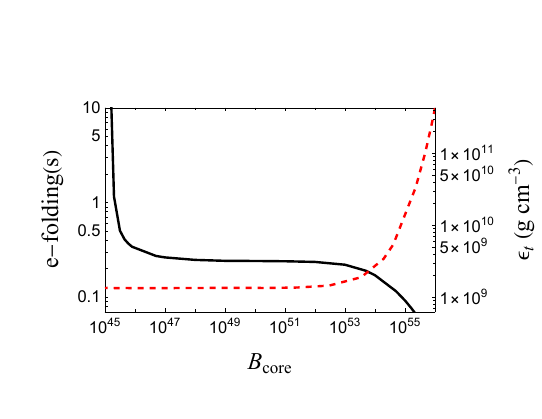}
    \caption{Dependence of the instability timescale on the core size. The solid black line shows the timescale of the mechanical instability as a function of the baryonic content of the core ($B_\mathrm{core})$. The dashed red line shows the density of nuclear matter at the interface with the SQM core. Plot from \citet{DiClemente:2022ktz}.}
    \label{collapse}
    \end{centering}
\end{figure}

When a WD gets close to Chandrasekhar's mass, large density fluctuations trigger nuclear fusion processes which end up with a supernova explosion. \citet{DiClemente:2022ktz} showed that, if the SQM core is large enough, the timescale of the gravitational collapse becomes much shorter than the timescale of nuclear reactions, leading the WD to a collapse instead of a supernova. 

\subsection{Central compact object in HESS J1731-347}
The study of the central compact object within the supernova remnant HESS J1731-347 indicates a small mass that is potentially subsolar \citep{Doroshenko2022}, namely an estimated mass of $0.7{7}_{-0.17}^{+0.20}\,\mathrm{M}_{\odot }$ and a radius $R=10.{4}_{-0.78}^{+0.86}$ km. Such attributes pose a conundrum for traditional models of supernova explosions, which typically fail to account for the existence of neutron stars with gravitational masses smaller than $\sim1.17$$\mathrm{M}_\odot$ \citep{Suwa:2018uni}. Hence, this compact object can be assumed to be a strange quark star \citep{DiClemente2022_hess}. 

To form a low-mass strange star, one can account for processes similar to those that result in the formation of light neutron stars. In the notable case of an electron-capture supernova, one has to consider a progenitor, which is a main-sequence star with a mass around 8$\mathrm{M}_\odot$ \citep{Suwa:2018uni}. For the progenitor to evolve into a strange star instead of a neutron star, it must have undergone an essential precondition: the accumulation of strange matter near its core. This accumulation could occur when, during the evolution of the star, the core is dense enough to stop a strangelet \citep{Madsen:1988zgf}. As the progenitor star undergoes the supernova event, the presence of accumulated strangelets favors the conversion of hadronic matter into strange matter, leading to the birth of a strange star with a relatively small mass.

Owing to the larger binding energy of strange quark stars, they can have a lower gravitational mass while having a baryonic mass equivalent to that of the lightest neutron stars ($\sim 1.28$$\mathrm{M}_\odot$). The strange star, just conserving the baryonic content, will have a mass that oscillates between 0.95$\mathrm{M}_\odot$ and 1.08$\mathrm{M}_\odot$, instead of a neutron star. It should be noted that the transition from hadronic to strange matter is markedly exothermic. Thus, during the supernova event, it is reasonable to hypothesize that a larger amount of matter is expelled, compared to the process of neutron star formation, thus resulting in the formation of a correspondingly lighter object.

\subsection{Solar flares}
In the analysis presented by \citet{BERTOLUCCI201713}, the correlation between solar phenomena such as flares and Extreme Ultraviolet (EUV) emissions and the heliocentric longitudes of Earth and the inner planets is explored, suggesting a potential link to increased solar activity during specific planetary alignments. This observation corroborates the hypothesis that slow-moving, galactic or cosmic-origin dark matter streams, potentially comprising anti-quark nuggets (a specific strangelets model), could be interacting with the Sun, influenced by the gravitational lensing effects of the planets. The peculiar behavior of these streams might be a significant contributor to various observed solar phenomena.

The study, while focused on a specific model of strangelets as proposed by \citet{Zhitnitsky:2002qa}, hints at a broader applicability. The observed correlations could be indicative of a class of dark matter that interacts more strongly than the traditionally considered WIMPs.

The underlying assumption is that this dark matter, regardless of its exact nature, interacts with the solar atmosphere in a way that could trigger increased solar activity. The solar atmosphere acts as a unique detector for these interactions due to its sensitivity to low-energy deposition, a condition not easily replicated with terrestrial dark matter detection methods. The potential for planetary gravitational lensing to focus these dark matter streams onto the Sun, thereby increasing the likelihood of solar flares or other activities, represents a novel approach to understand both solar dynamics and macro dark matter properties.

\section{Implications for cosmology}
Macroscopic dark matter candidates, like strangelets, exhibit a peculiar phenomenology related to the fact that, differently e.g. from WIMPs,  they interact with themselves and with ordinary matter. The effects of these interactions on cosmological observables are typically controlled by the reduced cross section $\sigma/M$. In this section, unless otherwise stated, we will assume that SQM makes up for all the dark matter content of the Universe.

Self-interacting dark matter has been proposed as an alternative to alleviate, or altogether eliminate, some shortcomings of the collisionless cold dark matter paradigm at galactic scales \citep{Carlson:1992fn, Spergel:1999mh}. Indeed, self-interacting dark matter, with velocity-independent reduced cross-section $\sigma_\mathrm{self}/M$ in the range $(0.1-1)\,\mathrm{cm}^2\,\mathrm{g}^{-1}$, would solve the core-cusp and ``too big to fail'' problems of cold dark matter, while at the same time satisfying constraints from massive galaxy clusters (see \citet{Bullock:2017xww} for a review). Notice anyway that the elastic reduced cross section is very small for clusters of SQM having the size discussed in the present paper:
\begin{equation}
   \frac{\sigma_\mathrm{self}}{M} = \frac{3}{\rho\,r_s} \sim 10^{-14} \left(\frac{r_s}{\mathrm{cm}}\right)^{-1}\,  \mathrm{cm^2\,g^{-1}} \sim 0.1 A^{-1/3} \mathrm{cm^2\,g^{-1}}  \, , \label{eq:cross_geom}
\end{equation}
where we have used the fact that $\sigma_\mathrm{self}$ is related to the geometrical cross section $\sigma_\mathrm{geom}$ through  $\sigma_\mathrm{self} = 4 \sigma_\mathrm{geom} = 4\pi r_s^2$. 

Interactions between dark matter particles and baryons generate a drag force that suppresses the growth of cosmic structures at small scales. This can be used to derive constraints on the dark matter-baryon scattering cross section. \citet{Dvorkin:2013cea} consider a velocity-dependent cross section $\sigma_\mathrm{scatt} \propto v^n$, with $-4 \le n \le 2$. For a velocity-independent cross section, they find $\sigma_\mathrm{scatt}/M < 3.3\times 10^{-3}\,\mathrm{cm^2}\,\mathrm{g}^{-1}$ at 95\% C.L. from a combination of CMB data from Planck and Ly-$\alpha$ data from the Sloan Digital Sky Survey. As in the case of dark matter self-interactions, these constraints are practically irrelevant for the clusters of SQM discussed in this paper.

Even though the elastic reduced cross section of strange matter is small, an open question concerns the inelastic cross section of this type of matter. In \citet{Bauswein:2009im} it has been shown that the "sticky properties" of strange matter, produced after the merger of two strange quark stars, deeply modify the dynamics of the system, if compared with that of ordinary matter. In particular, they lead to the formation of spiral arms which remain located close to the equatorial plane. The implications of this property for structure formation still need to be worked out. \citet{Jacobs:2014yca} argue that the constraints for inelastic self-interactions should be at least as strong as those for elastic scattering. Thus, in the absence of stronger arguments, one can treat the bounds on the elastic cross section as also being (conservative) bounds on the inelastic cross section. These are anyway automatically respected by SQM satisfying the bounds on the elastic cross section, since the absorption cross section is suppressed by the Coulomb barrier.

The electromagnetic properties of dark matter are also severely constrained by observations. The general argument that we have just sketched, about the drag force generated by dark matter-baryon scatterings, can be specified to the case of electromagnetic interactions. Constraints on the DM charge from Planck 2013 data were first derived by \citet{Dolgov:2013una}. \citet{Dvorkin:2013cea} express these constraints on the scattering cross section in terms of the DM charge ($\sigma_\mathrm{scatt} \propto v^{-4}$) and electric dipole moment ($\sigma_\mathrm{scatt} \propto v^{-2}$). In particular, the constraints on the DM charge\footnote{See also \citet{SinghSidhu:2019tbr} for a discussion of the conditions that allow to apply the constraints in \citet{Dvorkin:2013cea} to macroscopic dark matter candidates such as SQM.} $q$ read $|q|/e < 1.8 \times 10^{-6} (M/\mathrm{GeV})^{1/2}$. These constraints rely on imprints that would be left at the epoch of hydrogen recombination ($T \simeq 0.3\,\mathrm{eV})$ or later. SQM is charge neutral, thus to understand the relevance of this limits for the scenario considered in this paper we should study the ionization status of SQM at these times. A detailed study is beyond the scope of this paper, however we can gain some insight through a simple quantitative argument, reminiscent of the calculation of the recombination temperature of neutral hydrogen through the Saha equation. To this purpose, we make an educated guess about the energy $\Delta E_i$ needed to bring the SQM to the $i$-th from the $(i-1)$-th ionized state. The total ionization energy of ordinary atoms is of the order of $E_\mathrm{ion} \sim (16 \,\mathrm{eV}) \times Z^{7/3}$, so we can estimate $\Delta E_i \sim dE_\mathrm{ion}/dZ|_{Z=i} \simeq 30 \,\mathrm{eV} i^{4/3}$. The high density of photons with respect to SQM should also be taken into account, taking the role played by the larger (yet very small) baryon-to-photon ratio $\eta = n_b/n_\gamma \simeq 6\times 10^{-10}$ in estimates of the hydrogen recombination temperature. The quantity of relevance would thus be the ratio between the number densities of SQM and photons, $\eta_\mathrm{SQM} \equiv n_\mathrm{SQM}/n_\gamma \sim 5 (m_p/M) \eta $ where, for the sake of our order-of-magnitude estimates, we have neglected entropy productions taking place before hydrogen recombination. The factor $m_p/M \sim A^{-1}$ is potentially very small for the SQM clumps considered here; however, the dependence of the recombination temperature on $\eta$ (or $\eta_\mathrm{SQM}$) is only logarithmic. In fact, a direct numerical calculation from the Saha equation shows that, for $\eta_\mathrm{SQM}= 3\times 10^{-41}$ (corresponding to $A\sim 10^{32}$) at $T<10\,\mathrm{eV}$, the average ionization state of SQM should be $\lesssim 10$, i.e. each SQM particle behaves as charged dark matter with $|q| \lesssim 10 e$. Such a value of the charge is well below the limits reported in \citet{Dvorkin:2013cea}.

Finally, we mention the possibility that SQM might absorb standard model particles, in particular protons, neutrons and electrons. This scenario can be constrained by a number of early-time cosmological observables. For example, already \citet{Madsen:1985zx} noted that if SQM absorbs protons and neutrons at different rates (as expected in the case of a non-vanishing electrostatic potential) prior to the onset of Big Bang nucleosynthesis (BBN), the predicted abundances of light elements would be altered.
Following this line of reasoning, \citet{Jacobs:2014yca} have constrained the reduced cross-section as a function of the electrostatic potential, for generic macroscopic dark matter candidates. For positive values of the potential at the surface of the SQM of the order of $\sim 10\,\mathrm{MeV}$, 
as suggested in \citet{Alcock:1986hz}, they find a bound $\sigma_\mathrm{geom}/M \lesssim 2\times 10^{-10}\,\mathrm{cm}^2\,\mathrm{g}^{-1}$. For nuclear-density matter, this corresponds to $M\gtrsim 2\,\mathrm{g}$, which is satisfied by the SQM clumps considered here. \citet{Caloni:2021bwp} have instead considered the case in which the absorption starts after BBN; this scenario can be constrained by comparing the baryon abundances inferred from light element abundances on one side, and from the CMB anisotropies on the other. The bounds on the reduced cross section, and thus on the SQM mass, in this scenario are however looser than those reported in \citet{Jacobs:2014yca} for absorption taking place before BBN.

\section{Search for strangelets}

\begin{figure}
\begin{centering}
\includegraphics[width=\columnwidth]{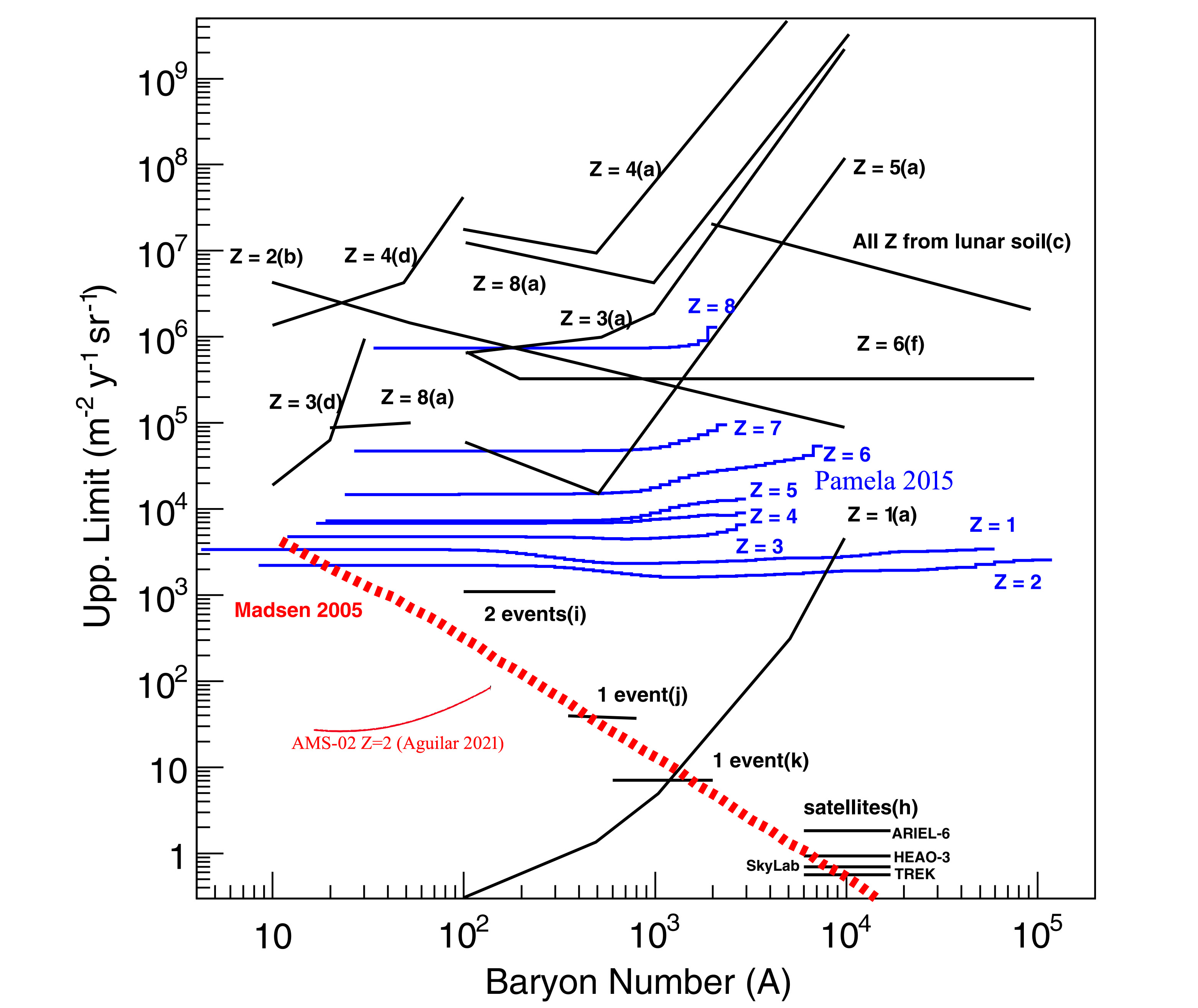}
    \caption{  Strangelet flux integral upper limits vs mass. The dotted line represents the dark matter constraint, assuming that all dark matter in the galaxy is composed by a single mass particle  \citep{Madsen:1988zgf, Madsen:1986jg}. The solid black lines represent 
experimental limits for strangelets converted into flux
limits. The curves labeled (a) \citep{PhysRevD.41.2074}, (b) \citep{PhysRevLett.92.022501}, (d) \citep{argonne}, (e) \citep{PhysRevLett.43.429}, and
(f) \citep{PhysRevD.30.1876} come from relic searches in terrestrial material. The
curve labeled (c) \citep{PhysRevLett.81.2416} comes from searches which bombard
materials with slow-moving heavy ions. The curves labeled with
(h) \citep{1987ApJ...314..739F, 1989ApJ...346..997B, 1978ApJ...220..719S, 1992Ap&SS.197..121P} represent satellite-based searches, obtained from the
exposure of the single experiments assuming Poissonian statistics. The curves labeled with (i)  \citep{PhysRevLett.65.2094}, (j) \citep{Ichimura1993}, and (k) \citep{PhysRevD.18.1382}
represent  detection of events consistent with strangelet
signals by balloon-borne cosmic-ray detectors. The blue curves are PAMELA measurements for nuclei up to $Z = 8$ \citet{PhysRevLett.115.111101}, the red one is the AMS-02 measurement for Z = 2 \citet{AGUILAR20211}.}
    \label{pamela}
    \end{centering}
\end{figure}

There have been several attempts to search  for strangelets or deeply penetrating particles in general. They employ different detection methodologies according to   the mass range and the velocity addressed.  
\begin{itemize}
   \item {\bf Light Mass ($A< 70$), non-moving particles}. SQM particles could have accumulated on the Moon or the Earth  surface and be detected with mass spectrometers. The Yale Wright Nuclear Structure Laboratory accelerator was employed to search for strangelet particles in lunar soil samples retrieved by the Apollo missions. This kind of  ground-based searches has the advantage of using a large amount of matter, providing high statistics, but assumes that strangelets can deposit on the upper layers of lunar or terrestrial soil. In \citet{PhysRevLett.103.092302} no events for Z = 5-9, and 11 with $42 < A < 70 $ were found, with a concentration of strangelets in lunar soil lower than $10^{-16}$ with respect to normal matter at 95\% C.L. 

     \item {\bf Light Mass ($A\lesssim 10^{5}$) , quasi-relativistic particles}. These particles can be directly detected in the flux of cosmic rays as they cross space-borne or balloon-borne detectors. 
In the 1990 balloon experiment HECRO-81 \citep{PhysRevLett.65.2094}  
reported the observation of two events with $Z \simeq 14$ and an estimated mass $  A \simeq 350$.
In 1987, the ARIEL-6 satellite (see \citet{1987ApJ...314..739F}) employing  Cherenkov counters, presented an analysis of $Z \geq 34$ finding no SQM candidates in 427 days of data acquisition.
 The HEAO-3 apparatus (see \citet{1989ApJ...346..997B}), with an exposure of
$8 \times 10^{11}$ $\mathrm{cm^2\: sr\: s}$,  did not find any strangelet  candidate.
Among the passive detectors (similar to the ground mica experiments) the SkyLab experiment \citep{1978ApJ...220..719S}, with 1.2 $\mathrm m^2$ Lexan
track detectors, did not find any valid candidate in the
superheavy ($Z > 110$) nuclei range.
On board the space station Mir, the experiment TREK \citep{Price1992} explored the $Z > 50$ region, also finding no strangelet candidate.
    In the  case of a magnetic spectrometer, strangelets would appear as events having a low deflection due to their high mass/charge (A/Z) ratio, due to the presence of the negative charge of the strange quarks (in case of Z=0 they would be undetectable). BESS balloon-borne detector flights  
have yielded no candidates for $5 \leq  Z \leq  26$ for $A/Z > 5$ \citep{Ichimura1993}. Subsequently, data  from 2006 to 2009 of the PAMELA space spectrometer were employed to conduct a direct search for SQM within cosmic rays \citep{Adriani2015}.   The spectrometer explored a rigidity range from $1-1000$ GV without finding traces of such particles, thereby setting an upper limit on the monochromatic strangelet flux in cosmic rays for particles having charge $1 \leq Z \leq 8$ and baryonic mass $4 \leq A \leq 1.2 \times 10^5$.  The AMS-01 Shuttle-Mir flight reported the observation of two events: $Z = 8$, $A = 20$ (3.93 GV) and $Z= 4$, $A = 50$ (5.13 GV), probably due to background, since  the following, much longer,  AMS-02 mission has found no events and  has lowered the limit to the flux of $Z=2$, yielding the limits represented in \autoref{pamela}.
\item {\bf Underground detector, Intermediate mass $10^{-11} \, \mathrm g < M \leq  1\,\mathrm g$. }
Underground, the MACRO detector placed a limit of  $1.1 \times  10^{-14}$ cm$^{-2}$ sr$^{-1}$ s$^{-1}$  and $  5.5 \times 10^{-15}$ cm$^{-2}$ sr$^{-1}$ s$^{-1}$in the  mass ranges 
$10^{-11} \,\mathrm g < m < 0.1 \,\mathrm g$ and $m > 0.1 \,\mathrm g $   respectively  \citep{AHLEN1991191},

  \item{ \bf High Mass, Meteor-like particle $10^{-6}   \,\mathrm g \lesssim M \lesssim 10^{4} \,\mathrm g$}.  Since strangelets can be rather massive, their presence can be inferred through the seismic activity they can potentially induce on the Moon, as suggested in \citet{BANERDT20061889}.  According to \citet{2006LPI....37.1048N}
  most of a rare class of lunar seismic events detected by the Apollo lunar seismic network occurred when the Moon faced a specific  direction of the celestial sphere, suggesting   that they may be caused by objects of  extra-solar origin. 
Strangelets can also give a meteor-like signal as they burn in the atmosphere, with tracks that are either longer and/or faster than those of conventional meteors.  Pi in the SKY,  searching for fast meteors in the sky \citep{LECH___PhysRevLett.125.091101}, yielded a limit close to the dark matter constraint. The DIMS experiment is currently searching for these events from the ground,  using two cameras for stereoscopic view of meteors and determination of their trajectory \citep{Abe:2021x8}. Similarly, Mini-EUSO is searching for these events from the ISS \citep{2015ExA....40..253A,Bacholle_2021} with a wider field. Another detector, SQM-ISS, aiming for direct detection of  SQM, is being designed to be placed on the ISS in the next years. In this case, the idea is to use a  local compact detector composed of a stack of scintillators (to detect ionization by passing particles) and a stack of metal plates (to read the vibration caused by the passage of particles) coupled to a  time-of-flight system capable of measuring the speed of the particle. 
Strangelets or similar particles are expected to move through the detector unaffected by the ISS hull, at speeds consistent with the galactic orbital velocity of $\sim$ 220 km/s or lower.
   
  \item {\bf High-Mass celestial bodies.} Asteroid-size objects having a mass $M\lesssim 10^{21} \,\mathrm g$, such as  Polyhymnia and Ludmilla \citep{2012P&SS...73...98C} both with mass densities of about $75$  g/cm$^3$ have been proposed as being constituted of either ultra-heavy (Z>120) nuclei or SQM. Although more prosaically their anomalous density is due to uncertainties in the mass determination related to the difficulty of orbital perturbations, a systematic search for anomalous density asteroids could place a limit in this mass range.  

  A systematic search for objects in the mass range $10^{17} \, \mathrm g \lesssim M \lesssim  10^{23}\,\mathrm g$ could be based on their gravitational effects on the orbits of planets or satellites. For instance, it has been shown in \citet{Tran:2023jci} that they can affect the orbit of Mars. Notice that the lower limit of $10^{17} \mathrm g$, indicated above, is due to the Hawking radiation as discussed in \autoref{subsec:comparisonPBH}. This limit applies to PBHs but not to strangelets which can have a smaller mass. Since direct observations of PBHs and strangelets are generally challenging, distinguishing between the two candidates can be difficult. However, detecting objects having a mass smaller than $10^{17} \, \mathrm g$ could offer a way to distinguish the two scenarios. It could be interesting to check if strangelets belonging to this mass range could be detected by an improved version of the Lunar Laser Ranging experiments \citep{Dickey:1994zz,Williams:2004qba}.
   
\end{itemize}

With a more refined theoretical framework to establish the possible sizes and eventually charges of strangelets, future experiments could be designed more effectively to search for these elusive particles. In particular, the space-based search, as opposed to ground-based one, can potentially identify strangelets without necessitating assumptions about their interaction with the Earth's atmosphere.

\section{Conclusions}
In this article we have revisited the possibility that dark matter is made of clusters of up, down and strange quarks, as suggested by \citet{Witten:1984rs}. We have shown that it is at least conceptually possible to formulate a scheme, leading to the formation of those objects in the Early Universe, without assuming the existence of a first-order phase transition associated with confinement. 
In our analysis we have considered the possibility of producing massive strangelets through a first-order phase transition taking place in the high-temperature hadronic matter. A first possibility is based on the coalescence of sexaquarks \citep{Shahrbaf:2022upc}. Another possibility is instead related to the existence of a chemical-mechanical instability leading to the production of high-density, finite-strangeness matter \citep{Lavagno:2022orw}. These theoretical possibilities need to be further investigated. We have then discussed how strangelets undergo a partial evaporation, and we have shown that these objects can satisfy all the astrophysical and cosmological constraints on dark matter. This result clearly depends on the value of a few phenomenological parameters, for instance the one related to the effect of Pauli blocking on the evaporation rate. It will be important to investigate if the value of these parameters can be justified by microphysics.

We have shown that SQM is not strongly self-interacting and that it can be considered charge neutral. On the other hand, it has a finite inelastic cross section with itself and with ordinary matter, but the implications of this property for the evolution of galaxies still need to be worked out.

\section*{Acknowledgements}

This material is based upon work supported by the National Science Foundation under grants No. PHY-2208724,
PHY-1654219 and PHY-2116686, and within the framework
of the MUSES collaboration, under grant number No. OAC-
2103680. This material is also based upon work supported
by the U.S. Department of Energy, Office of Science, Office of Nuclear Physics, under Award Number DE-SC0022023 as well as by the National Aeronautics and Space Agency (NASA) under Award Number 80NSSC24K0767.

\section*{Data Availability}

The data underlying this article will be shared on reasonable request to the corresponding author.

\bibliographystyle{mnras}
\bibliography{bibliography,casol_bib} 

\begin{thebibliography}{}
\makeatletter
\relax
\def\mn@urlcharsother{\let\do\@makeother \do\$\do\&\do\#\do\^\do\_\do\%\do\~}
\def\mn@doi{\begingroup\mn@urlcharsother \@ifnextchar [ {\mn@doi@} {\mn@doi@[]}}
\def\mn@doi@[#1]#2{\def\@tempa{#1}\ifx\@tempa\@empty \href {http://dx.doi.org/#2} {doi:#2}\else \href {http://dx.doi.org/#2} {#1}\fi \endgroup}
\def\mn@eprint#1#2{\mn@eprint@#1:#2::\@nil}
\def\mn@eprint@arXiv#1{\href {http://arxiv.org/abs/#1} {{\tt arXiv:#1}}}
\def\mn@eprint@dblp#1{\href {http://dblp.uni-trier.de/rec/bibtex/#1.xml} {dblp:#1}}
\def\mn@eprint@#1:#2:#3:#4\@nil{\def\@tempa {#1}\def\@tempb {#2}\def\@tempc {#3}\ifx \@tempc \@empty \let \@tempc \@tempb \let \@tempb \@tempa \fi \ifx \@tempb \@empty \def\@tempb {arXiv}\fi \@ifundefined {mn@eprint@\@tempb}{\@tempb:\@tempc}{\expandafter \expandafter \csname mn@eprint@\@tempb\endcsname \expandafter{\@tempc}}}

\bibitem[\protect\citeauthoryear{Aaij et~al.}{Aaij et~al.}{2024}]{LHCb:2023wbo}
Aaij R.,  et~al., 2024, \mn@doi [Phys. Rev. Lett.] {10.1103/PhysRevLett.132.081901}, 132, 081901

\bibitem[\protect\citeauthoryear{Abe et~al.,}{Abe et~al.}{2021}]{Abe:2021x8}
Abe S.,  et~al., 2021, \mn@doi [PoS] {10.22323/1.395.0554}, ICRC2021, 554

\bibitem[\protect\citeauthoryear{Adam et~al.}{Adam et~al.}{2016}]{ALICE:2015oer}
Adam J.,  et~al., 2016, \mn@doi [Phys. Lett. B] {10.1016/j.physletb.2016.01.040}, 754, 360

\bibitem[\protect\citeauthoryear{{Adams} et~al.,}{{Adams} et~al.}{2015}]{2015ExA....40..253A}
{Adams} J.~H.,  et~al., 2015, \mn@doi [Experimental Astronomy] {10.1007/s10686-014-9375-4}, \href {http://adsabs.harvard.edu/abs/2015ExA....40..253A} {40, 253}

\bibitem[\protect\citeauthoryear{Adler \& Piran}{Adler \& Piran}{1984}]{Adler:1983zh}
Adler S.~L.,  Piran T.,  1984, \mn@doi [Rev. Mod. Phys.] {10.1103/RevModPhys.56.1}, 56, 1

\bibitem[\protect\citeauthoryear{Adriani et~al.,}{Adriani et~al.}{2015a}]{Adriani2015}
Adriani O.,  et~al., 2015a, \mn@doi [Phys. Rev. Lett.] {10.1103/PhysRevLett.115.111101}, 115, 111101

\bibitem[\protect\citeauthoryear{Adriani et~al.,}{Adriani et~al.}{2015b}]{PhysRevLett.115.111101}
Adriani O.,  et~al., 2015b, \mn@doi [Phys. Rev. Lett.] {10.1103/PhysRevLett.115.111101}, 115, 111101

\bibitem[\protect\citeauthoryear{Afzal et~al.,}{Afzal et~al.}{2023}]{Afzal_2023}
Afzal A.,  et~al., 2023, \mn@doi [The Astrophysical Journal Letters] {10.3847/2041-8213/acdc91}, 951, L11

\bibitem[\protect\citeauthoryear{Aguilar et~al.,}{Aguilar et~al.}{2021}]{AGUILAR20211}
Aguilar M.,  et~al., 2021, \mn@doi [Physics Reports] {https://doi.org/10.1016/j.physrep.2020.09.003}, 894, 1

\bibitem[\protect\citeauthoryear{Ahlen et~al.,}{Ahlen et~al.}{1991}]{AHLEN1991191}
Ahlen S.,  et~al., 1991, \mn@doi [Nuclear Physics B - Proceedings Supplements] {https://doi.org/10.1016/0920-5632(91)90323-7}, 24, 191

\bibitem[\protect\citeauthoryear{Alba, Sarti, Noronha-Hostler, Parotto, Portillo-Vazquez, Ratti  \& Stafford}{Alba et~al.}{2020}]{Alba:2020jir}
Alba P.,  Sarti V.~M.,  Noronha-Hostler J.,  Parotto P.,  Portillo-Vazquez I.,  Ratti C.,   Stafford J.~M.,  2020, \mn@doi [Phys. Rev. C] {10.1103/PhysRevC.101.054905}, 101, 054905

\bibitem[\protect\citeauthoryear{Alberico, Drago  \& Ratti}{Alberico et~al.}{2002}]{Alberico:2001zz}
Alberico W.~M.,  Drago A.,   Ratti C.,  2002, \mn@doi [Nucl. Phys. A] {10.1016/S0375-9474(02)00680-2}, 706, 143

\bibitem[\protect\citeauthoryear{Alcock \& Farhi}{Alcock \& Farhi}{1985}]{Alcock:1985vc}
Alcock C.,  Farhi E.,  1985, \mn@doi [Phys. Rev. D] {10.1103/PhysRevD.32.1273}, 32, 1273

\bibitem[\protect\citeauthoryear{Alcock, Farhi  \& Olinto}{Alcock et~al.}{1986}]{Alcock:1986hz}
Alcock C.,  Farhi E.,   Olinto A.,  1986, \mn@doi [Astrophys. J.] {10.1086/164679}, 310, 261

\bibitem[\protect\citeauthoryear{Alcock et~al.,}{Alcock et~al.}{2000}]{Alcock:2000ph}
Alcock C.,  et~al., 2000, \mn@doi [The Astrophysical Journal] {10.1086/309512}, 542, 281

\bibitem[\protect\citeauthoryear{Alford \& Reddy}{Alford \& Reddy}{2003}]{Alford:2002rj}
Alford M.,  Reddy S.,  2003, \mn@doi [Phys. Rev. D] {10.1103/PhysRevD.67.074024}, 67, 074024

\bibitem[\protect\citeauthoryear{Alford, Berges  \& Rajagopal}{Alford et~al.}{1999}]{Alford_1999}
Alford M.,  Berges J.,   Rajagopal K.,  1999, \mn@doi [Nuclear Physics B] {10.1016/s0550-3213(99)00410-1}, 558, 219–242

\bibitem[\protect\citeauthoryear{Alford, Schmitt, Rajagopal  \& Sch\"afer}{Alford et~al.}{2008}]{Alford:2007xm}
Alford M.~G.,  Schmitt A.,  Rajagopal K.,   Sch\"afer T.,  2008, \mn@doi [Rev. Mod. Phys.] {10.1103/RevModPhys.80.1455}, 80, 1455

\bibitem[\protect\citeauthoryear{Alford, Harris  \& Sachdeva}{Alford et~al.}{2017}]{Alford:2017vca}
Alford M.~G.,  Harris S.~P.,   Sachdeva P.~S.,  2017, \mn@doi [Astrophys. J.] {10.3847/1538-4357/aa8509}, 847, 109

\bibitem[\protect\citeauthoryear{Amore, Birse, McGovern  \& Walet}{Amore et~al.}{2002}]{Amore:2001uf}
Amore P.,  Birse M.~C.,  McGovern J.~A.,   Walet N.~R.,  2002, \mn@doi [Phys. Rev. D] {10.1103/PhysRevD.65.074005}, 65, 074005

\bibitem[\protect\citeauthoryear{Anglani, Casalbuoni, Ciminale, Ippolito, Gatto, Mannarelli  \& Ruggieri}{Anglani et~al.}{2014}]{Anglani:2013gfu}
Anglani R.,  Casalbuoni R.,  Ciminale M.,  Ippolito N.,  Gatto R.,  Mannarelli M.,   Ruggieri M.,  2014, \mn@doi [Rev. Mod. Phys.] {10.1103/RevModPhys.86.509}, 86, 509

\bibitem[\protect\citeauthoryear{Aoki, Endrodi, Fodor, Katz  \& Szabo}{Aoki et~al.}{2006}]{Aoki:2006we}
Aoki Y.,  Endrodi G.,  Fodor Z.,  Katz S.~D.,   Szabo K.~K.,  2006, \mn@doi [Nature] {10.1038/nature05120}, 443, 675

\bibitem[\protect\citeauthoryear{Bacholle et~al.,}{Bacholle et~al.}{2021}]{Bacholle_2021}
Bacholle S.,  et~al., 2021, \mn@doi [The Astrophysical Journal Supplement Series] {10.3847/1538-4365/abd93d}, 253, 36

\bibitem[\protect\citeauthoryear{Bai, Chen  \& Korwar}{Bai et~al.}{2023}]{Bai:2023cqj}
Bai Y.,  Chen T.-K.,   Korwar M.,  2023, \mn@doi [JHEP] {10.1007/JHEP12(2023)194}, 12, 194

\bibitem[\protect\citeauthoryear{Banerdt, Chui, Herrin, Rosenbaum  \& Teplitz}{Banerdt et~al.}{2006}]{BANERDT20061889}
Banerdt W.~B.,  Chui T.,  Herrin E.~T.,  Rosenbaum D.,   Teplitz V.~L.,  2006, \mn@doi [Advances in Space Research] {https://doi.org/10.1016/j.asr.2005.06.034}, 37, 1889

\bibitem[\protect\citeauthoryear{Barnacka, Glicenstein  \& Moderski}{Barnacka et~al.}{2012}]{Barnacka_2012}
Barnacka A.,  Glicenstein J.-F.,   Moderski R.,  2012, \mn@doi [Physical Review D] {10.1103/physrevd.86.043001}, 86

\bibitem[\protect\citeauthoryear{Bauswein, Oechslin  \& Janka}{Bauswein et~al.}{2010}]{Bauswein:2009im}
Bauswein A.,  Oechslin R.,   Janka H.~T.,  2010, \mn@doi [Phys. Rev. D] {10.1103/PhysRevD.81.024012}, 81, 024012

\bibitem[\protect\citeauthoryear{Bellwied, Borsanyi, Fodor, Katz  \& Ratti}{Bellwied et~al.}{2013}]{Bellwied:2013cta}
Bellwied R.,  Borsanyi S.,  Fodor Z.,  Katz S.~D.,   Ratti C.,  2013, \mn@doi [Phys. Rev. Lett.] {10.1103/PhysRevLett.111.202302}, 111, 202302

\bibitem[\protect\citeauthoryear{Bellwied et~al.,}{Bellwied et~al.}{2021}]{Bellwied:2021nrt}
Bellwied R.,  et~al., 2021, \mn@doi [Phys. Rev. D] {10.1103/PhysRevD.104.094508}, 104, 094508

\bibitem[\protect\citeauthoryear{Bertolucci, Zioutas, Hofmann  \& Maroudas}{Bertolucci et~al.}{2017}]{BERTOLUCCI201713}
Bertolucci S.,  Zioutas K.,  Hofmann S.,   Maroudas M.,  2017, \mn@doi [Physics of the Dark Universe] {https://doi.org/10.1016/j.dark.2017.06.001}, 17, 13

\bibitem[\protect\citeauthoryear{{Binns} et~al.,}{{Binns} et~al.}{1989}]{1989ApJ...346..997B}
{Binns} W.~R.,  et~al., 1989, \mn@doi [\apj] {10.1086/168082}, \href {https://ui.adsabs.harvard.edu/abs/1989ApJ...346..997B} {346, 997}

\bibitem[\protect\citeauthoryear{Bodmer}{Bodmer}{1971}]{Bodmer:1971we}
Bodmer A.~R.,  1971, \mn@doi [Phys. Rev. D] {10.1103/PhysRevD.4.1601}, 4, 1601

\bibitem[\protect\citeauthoryear{Brandt, Endrodi  \& Schmalzbauer}{Brandt et~al.}{2018}]{Brandt:2017oyy}
Brandt B.~B.,  Endrodi G.,   Schmalzbauer S.,  2018, \mn@doi [Phys. Rev. D] {10.1103/PhysRevD.97.054514}, 97, 054514

\bibitem[\protect\citeauthoryear{Braun-Munzinger \& D\"onigus}{Braun-Munzinger \& D\"onigus}{2019}]{Braun-Munzinger:2018hat}
Braun-Munzinger P.,  D\"onigus B.,  2019, \mn@doi [Nucl. Phys. A] {10.1016/j.nuclphysa.2019.02.006}, 987, 144

\bibitem[\protect\citeauthoryear{Buballa}{Buballa}{2005}]{Buballa:2003qv}
Buballa M.,  2005, \mn@doi [Phys. Rept.] {10.1016/j.physrep.2004.11.004}, 407, 205

\bibitem[\protect\citeauthoryear{Bucciantini, Drago, Pagliara, Traversi  \& Bauswein}{Bucciantini et~al.}{2022}]{Bucciantini:2019ivq}
Bucciantini N.,  Drago A.,  Pagliara G.,  Traversi S.,   Bauswein A.,  2022, \mn@doi [Phys. Rev. D] {10.1103/PhysRevD.106.103032}, 106, 103032

\bibitem[\protect\citeauthoryear{Bullock \& Boylan-Kolchin}{Bullock \& Boylan-Kolchin}{2017}]{Bullock:2017xww}
Bullock J.~S.,  Boylan-Kolchin M.,  2017, \mn@doi [Ann. Rev. Astron. Astrophys.] {10.1146/annurev-astro-091916-055313}, 55, 343

\bibitem[\protect\citeauthoryear{Burdin, Fairbairn, Mermod, Milstead, Pinfold, Sloan  \& Taylor}{Burdin et~al.}{2015}]{Burdin:2014xma}
Burdin S.,  Fairbairn M.,  Mermod P.,  Milstead D.,  Pinfold J.,  Sloan T.,   Taylor W.,  2015, \mn@doi [Phys. Rept.] {10.1016/j.physrep.2015.03.004}, 582, 1

\bibitem[\protect\citeauthoryear{Caloni, Gerbino  \& Lattanzi}{Caloni et~al.}{2021}]{Caloni:2021bwp}
Caloni L.,  Gerbino M.,   Lattanzi M.,  2021, \mn@doi [JCAP] {10.1088/1475-7516/2021/07/027}, 07, 027

\bibitem[\protect\citeauthoryear{Carlson, Machacek  \& Hall}{Carlson et~al.}{1992}]{Carlson:1992fn}
Carlson E.~D.,  Machacek M.~E.,   Hall L.~J.,  1992, \mn@doi [Astrophys. J.] {10.1086/171833}, 398, 43

\bibitem[\protect\citeauthoryear{Carr}{Carr}{1975}]{Carr:1975qj}
Carr B.~J.,  1975, \mn@doi [Astrophys. J.] {10.1086/153853}, 201, 1

\bibitem[\protect\citeauthoryear{Carr \& Hawking}{Carr \& Hawking}{1974}]{Carr:1974nx}
Carr B.~J.,  Hawking S.~W.,  1974, \mn@doi [Mon. Not. Roy. Astron. Soc.] {10.1093/mnras/168.2.399}, 168, 399

\bibitem[\protect\citeauthoryear{Carr, Kohri, Sendouda  \& Yokoyama}{Carr et~al.}{2021}]{Carr:2020gox}
Carr B.,  Kohri K.,  Sendouda Y.,   Yokoyama J.,  2021, \mn@doi [Rept. Prog. Phys.] {10.1088/1361-6633/ac1e31}, 84, 116902

\bibitem[\protect\citeauthoryear{{Carry}}{{Carry}}{2012}]{2012P&SS...73...98C}
{Carry} B.,  2012, \mn@doi [\planss] {10.1016/j.pss.2012.03.009}, \href {https://ui.adsabs.harvard.edu/abs/2012P&SS...73...98C} {73, 98}

\bibitem[\protect\citeauthoryear{Chodos, Jaffe, Johnson, Thorn  \& Weisskopf}{Chodos et~al.}{1974}]{Chodos1974}
Chodos A.,  Jaffe R.~L.,  Johnson K.,  Thorn C.~B.,   Weisskopf V.~F.,  1974, Physical Review D, 9, 3471

\bibitem[\protect\citeauthoryear{Christensen \& Moloney}{Christensen \& Moloney}{2005}]{christensen2005complexity}
Christensen K.,  Moloney N.,  2005, Complexity and Criticality.
Advanced physics texts, Imperial College Press, \url {https://books.google.it/books?id=bAIM1_EoQu0C}

\bibitem[\protect\citeauthoryear{Cyrol, Pawlowski, Rothkopf  \& Wink}{Cyrol et~al.}{2018}]{Cyrol_2018}
Cyrol A.~K.,  Pawlowski J.~M.,  Rothkopf A.,   Wink N.,  2018, \mn@doi [SciPost Physics] {10.21468/scipostphys.5.6.065}, 5

\bibitem[\protect\citeauthoryear{Davis \& Smith}{Davis \& Smith}{1973}]{Davis1973}
Davis S.,  Smith A.,  1973, \mn@doi [Kolloid-Zeitschrift und Zeitschrift für Polymere] {10.1007/BF01498732}

\bibitem[\protect\citeauthoryear{Di~Clemente, Mannarelli  \& Tonelli}{Di~Clemente et~al.}{2020}]{DiClemente:2020szl}
Di~Clemente F.,  Mannarelli M.,   Tonelli F.,  2020, \mn@doi [Phys. Rev. D] {10.1103/PhysRevD.101.103003}, 101, 103003

\bibitem[\protect\citeauthoryear{{Di Clemente}, {Drago}  \& {Pagliara}}{{Di Clemente} et~al.}{2022}]{DiClemente2022_hess}
{Di Clemente} F.,  {Drago} A.,   {Pagliara} G.,  2022, \mn@doi [arXiv e-prints] {10.48550/arXiv.2211.07485}, \href {https://ui.adsabs.harvard.edu/abs/2022arXiv221107485D} {p. arXiv:2211.07485}

\bibitem[\protect\citeauthoryear{Di~Clemente, Drago, Char  \& Pagliara}{Di~Clemente et~al.}{2023}]{DiClemente:2022ktz}
Di~Clemente F.,  Drago A.,  Char P.,   Pagliara G.,  2023, \mn@doi [Astron. Astrophys.] {10.1051/0004-6361/202347607}, 678, L1

\bibitem[\protect\citeauthoryear{Di~Salvo, Sanna, Burderi, Papitto, Iaria, Gambino  \& Riggio}{Di~Salvo et~al.}{2019}]{DiSalvo:2018mua}
Di~Salvo T.,  Sanna A.,  Burderi L.,  Papitto A.,  Iaria R.,  Gambino A.~F.,   Riggio A.,  2019, \mn@doi [Mon. Not. Roy. Astron. Soc.] {10.1093/mnras/sty2974}, 483, 767

\bibitem[\protect\citeauthoryear{Dickey et~al.}{Dickey et~al.}{1994}]{Dickey:1994zz}
Dickey J.~O.,  et~al., 1994, \mn@doi [Science] {10.1126/science.265.5171.482}, 265, 482

\bibitem[\protect\citeauthoryear{Dolgov, Dubovsky, Rubtsov  \& Tkachev}{Dolgov et~al.}{2013}]{Dolgov:2013una}
Dolgov A.~D.,  Dubovsky S.~L.,  Rubtsov G.~I.,   Tkachev I.~I.,  2013, \mn@doi [Phys. Rev. D] {10.1103/PhysRevD.88.117701}, 88, 117701

\bibitem[\protect\citeauthoryear{Dolgov, Kuranov, Mitichkin, Porey, Postnov, Sazhina  \& Simkin}{Dolgov et~al.}{2020}]{Dolgov_2020}
Dolgov A.,  Kuranov A.,  Mitichkin N.,  Porey S.,  Postnov K.,  Sazhina O.,   Simkin I.,  2020, \mn@doi [Journal of Cosmology and Astroparticle Physics] {10.1088/1475-7516/2020/12/017}, 2020, 017–017

\bibitem[\protect\citeauthoryear{Dondi, Drago  \& Pagliara}{Dondi et~al.}{2017}]{Dondi_2017}
Dondi N.,  Drago A.,   Pagliara G.,  2017, \mn@doi [{EPJ} Web of Conferences] {10.1051/epjconf/201713709004}, 137, 09004

\bibitem[\protect\citeauthoryear{Doroshenko, Suleimanov, Puehlhofer  \& Santangelo}{Doroshenko et~al.}{2022}]{Doroshenko2022}
Doroshenko V.,  Suleimanov V.,  Puehlhofer G.,   Santangelo A.,  2022, \mn@doi [Nature Astronomy] {10.1038/s41550-022-01800-1}

\bibitem[\protect\citeauthoryear{Drago \& Lavagno}{Drago \& Lavagno}{2001}]{Drago:2001nq}
Drago A.,  Lavagno A.,  2001, \mn@doi [Phys. Lett. B] {10.1016/S0370-2693(01)00579-2}, 511, 229

\bibitem[\protect\citeauthoryear{Drago \& Pagliara}{Drago \& Pagliara}{2016}]{Drago:2015dea}
Drago A.,  Pagliara G.,  2016, \mn@doi [Eur. Phys. J. A] {10.1140/epja/i2016-16041-2}, 52, 41

\bibitem[\protect\citeauthoryear{Drago, Lavagno  \& Pagliara}{Drago et~al.}{2004}]{Drago:2004vu}
Drago A.,  Lavagno A.,   Pagliara G.,  2004, \mn@doi [Phys. Rev. D] {10.1103/PhysRevD.69.057505}, 69, 057505

\bibitem[\protect\citeauthoryear{Drago, Lavagno  \& Pagliara}{Drago et~al.}{2014a}]{Drago:2013fsa}
Drago A.,  Lavagno A.,   Pagliara G.,  2014a, \mn@doi [Phys. Rev. D] {10.1103/PhysRevD.89.043014}, 89, 043014

\bibitem[\protect\citeauthoryear{Drago, Lavagno, Pagliara  \& Pigato}{Drago et~al.}{2014b}]{DragoLavagnoHyperons2014}
Drago A.,  Lavagno A.,  Pagliara G.,   Pigato D.,  2014b, \mn@doi [Phys. Rev. C] {10.1103/PhysRevC.90.065809}, 90, 065809

\bibitem[\protect\citeauthoryear{Drago, Lavagno, Pagliara  \& Pigato}{Drago et~al.}{2016}]{Drago:2015cea}
Drago A.,  Lavagno A.,  Pagliara G.,   Pigato D.,  2016, \mn@doi [Eur. Phys. J. A] {10.1140/epja/i2016-16040-3}, 52, 40

\bibitem[\protect\citeauthoryear{Dvorkin, Blum  \& Kamionkowski}{Dvorkin et~al.}{2014}]{Dvorkin:2013cea}
Dvorkin C.,  Blum K.,   Kamionkowski M.,  2014, \mn@doi [Phys. Rev. D] {10.1103/PhysRevD.89.023519}, 89, 023519

\bibitem[\protect\citeauthoryear{{EPTA Collaboration} et~al.,}{{EPTA Collaboration} et~al.}{2024}]{EPTA_2024}
{EPTA Collaboration} et~al., 2024, \mn@doi [\aap] {10.1051/0004-6361/202347433}, \href {https://ui.adsabs.harvard.edu/abs/2024A&A...685A..94E} {685, A94}

\bibitem[\protect\citeauthoryear{Farhi \& Jaffe}{Farhi \& Jaffe}{1984}]{Farhi:1984qu}
Farhi E.,  Jaffe R.~L.,  1984, \mn@doi [Phys. Rev. D] {10.1103/PhysRevD.30.2379}, 30, 2379

\bibitem[\protect\citeauthoryear{Farrar}{Farrar}{2022}]{Farrar:2022mih}
Farrar G.~R.,  2022, {A Stable Sexaquark: Overview and Discovery Strategies} (\mn@eprint {arXiv} {2201.01334})

\bibitem[\protect\citeauthoryear{Flor, Olinger  \& Bellwied}{Flor et~al.}{2021}]{Flor:2020fdw}
Flor F.~A.,  Olinger G.,   Bellwied R.,  2021, \mn@doi [Phys. Lett. B] {10.1016/j.physletb.2021.136098}, 814, 136098

\bibitem[\protect\citeauthoryear{{Fowler}, {Walker}, {Masheder}, {Moses}, {Worley}  \& {Gay}}{{Fowler} et~al.}{1987}]{1987ApJ...314..739F}
{Fowler} P.~H.,  {Walker} R.~N.~F.,  {Masheder} M.~R.~W.,  {Moses} R.~T.,  {Worley} A.,   {Gay} A.~M.,  1987, \mn@doi [\apj] {10.1086/165101}, \href {https://ui.adsabs.harvard.edu/abs/1987ApJ...314..739F} {314, 739}

\bibitem[\protect\citeauthoryear{Freedman \& McLerran}{Freedman \& McLerran}{1978}]{Freedman:1977gz}
Freedman B.,  McLerran L.~D.,  1978, \mn@doi [Phys. Rev. D] {10.1103/PhysRevD.17.1109}, 17, 1109

\bibitem[\protect\citeauthoryear{Froustey \& Pitrou}{Froustey \& Pitrou}{2024}]{Froustey_2024}
Froustey J.,  Pitrou C.,  2024, \mn@doi [Physical Review D] {10.1103/physrevd.110.103551}, 110

\bibitem[\protect\citeauthoryear{Gao \& Oldengott}{Gao \& Oldengott}{2022}]{Gao:2021nwz}
Gao F.,  Oldengott I.~M.,  2022, \mn@doi [Phys. Rev. Lett.] {10.1103/PhysRevLett.128.131301}, 128, 131301

\bibitem[\protect\citeauthoryear{Glendenning, Kettner  \& Weber}{Glendenning et~al.}{1995a}]{Glendenning:1994sp}
Glendenning N.~K.,  Kettner C.,   Weber F.,  1995a, \mn@doi [Phys. Rev. Lett.] {10.1103/PhysRevLett.74.3519}, 74, 3519

\bibitem[\protect\citeauthoryear{Glendenning, Kettner  \& Weber}{Glendenning et~al.}{1995b}]{Glendenning:1994zb}
Glendenning N.~K.,  Kettner C.,   Weber F.,  1995b, \mn@doi [Astrophys. J.] {10.1086/176136}, 450, 253

\bibitem[\protect\citeauthoryear{{Haensel}, {Zdunik}  \& {Schaefer}}{{Haensel} et~al.}{1986}]{Haensel:1986}
{Haensel} P.,  {Zdunik} J.~L.,   {Schaefer} R.,  1986, \aap, \href {https://ui.adsabs.harvard.edu/abs/1986A&A...160..121H} {160, 121}

\bibitem[\protect\citeauthoryear{Han et~al.,}{Han et~al.}{2009}]{PhysRevLett.103.092302}
Han K.,  et~al., 2009, \mn@doi [Phys. Rev. Lett.] {10.1103/PhysRevLett.103.092302}, 103, 092302

\bibitem[\protect\citeauthoryear{Hawking}{Hawking}{1974}]{Hawking:1974rv}
Hawking S.~W.,  1974, \mn@doi [Nature] {10.1038/248030a0}, 248, 30

\bibitem[\protect\citeauthoryear{Hawking}{Hawking}{1975}]{Hawking:1975vcx}
Hawking S.~W.,  1975, \mn@doi [Commun. Math. Phys.] {10.1007/BF02345020}, 43, 199

\bibitem[\protect\citeauthoryear{Hemmick et~al.,}{Hemmick et~al.}{1990}]{PhysRevD.41.2074}
Hemmick T.~K.,  et~al., 1990, \mn@doi [Phys. Rev. D] {10.1103/PhysRevD.41.2074}, 41, 2074

\bibitem[\protect\citeauthoryear{Herrin, Rosenbaum  \& Teplitz}{Herrin et~al.}{2006}]{Herrin:2005kb}
Herrin E.~T.,  Rosenbaum D.~C.,   Teplitz V.~L.,  2006, \mn@doi [Phys. Rev. D] {10.1103/PhysRevD.73.043511}, 73, 043511

\bibitem[\protect\citeauthoryear{Horak, Ihssen, Papavassiliou, Pawlowski, Weber  \& Wetterich}{Horak et~al.}{2022}]{Horak:2022aqx}
Horak J.,  Ihssen F.,  Papavassiliou J.,  Pawlowski J.~M.,  Weber A.,   Wetterich C.,  2022, \mn@doi [SciPost Phys.] {10.21468/SciPostPhys.13.2.042}, 13, 042

\bibitem[\protect\citeauthoryear{Huber}{Huber}{2020}]{HUBER20201}
Huber M.~Q.,  2020, \mn@doi [Physics Reports] {https://doi.org/10.1016/j.physrep.2020.04.004}, 879, 1

\bibitem[\protect\citeauthoryear{Ichimura et~al.,}{Ichimura et~al.}{1993}]{Ichimura1993}
Ichimura M.,  et~al., 1993, \mn@doi [Il Nuovo Cimento A (1965-1970)] {10.1007/BF02771498}, 106, 843

\bibitem[\protect\citeauthoryear{Isaac et~al.,}{Isaac et~al.}{1998}]{PhysRevLett.81.2416}
Isaac M. C.~P.,  et~al., 1998, \mn@doi [Phys. Rev. Lett.] {10.1103/PhysRevLett.81.2416}, 81, 2416

\bibitem[\protect\citeauthoryear{Jacobs, Starkman  \& Lynn}{Jacobs et~al.}{2015}]{Jacobs:2014yca}
Jacobs D.~M.,  Starkman G.~D.,   Lynn B.~W.,  2015, \mn@doi [Mon. Not. Roy. Astron. Soc.] {10.1093/mnras/stv774}, 450, 3418

\bibitem[\protect\citeauthoryear{Jaffe}{Jaffe}{1977}]{Jaffe1977}
Jaffe R.~L.,  1977, \mn@doi [Phys. Rev. Lett.] {10.1103/PhysRevLett.38.195}, 38, 195

\bibitem[\protect\citeauthoryear{Katz, Kopp, Sibiryakov  \& Xue}{Katz et~al.}{2018}]{Katz:2018zrn}
Katz A.,  Kopp J.,  Sibiryakov S.,   Xue W.,  2018, \mn@doi [JCAP] {10.1088/1475-7516/2018/12/005}, 12, 005

\bibitem[\protect\citeauthoryear{Kettner, Weber, Weigel  \& Glendenning}{Kettner et~al.}{1995}]{Kettner:1994zs}
Kettner C.,  Weber F.,  Weigel M.~K.,   Glendenning N.~K.,  1995, \mn@doi [Phys. Rev. D] {10.1103/PhysRevD.51.1440}, 51, 1440

\bibitem[\protect\citeauthoryear{Klahn \& Fischer}{Klahn \& Fischer}{2015}]{Klahn:2015mfa}
Klahn T.,  Fischer T.,  2015, \mn@doi [Astrophys. J.] {10.1088/0004-637X/810/2/134}, 810, 134

\bibitem[\protect\citeauthoryear{Klein, Midleton  \& Stephens}{Klein et~al.}{}]{argonne}
Klein J.,  Midleton R.,   Stephens W.~E., , Argonne National Laboratory Report No. ANL/PHY-81-1, 1981

\bibitem[\protect\citeauthoryear{Kurban, Huang, Geng  \& Zong}{Kurban et~al.}{2022}]{Kurban:2020xtb}
Kurban A.,  Huang Y.-F.,  Geng J.-J.,   Zong H.-S.,  2022, \mn@doi [Phys. Lett. B] {10.1016/j.physletb.2022.137204}, 832, 137204

\bibitem[\protect\citeauthoryear{Kurkela, Romatschke  \& Vuorinen}{Kurkela et~al.}{2010}]{Kurkela:2009gj}
Kurkela A.,  Romatschke P.,   Vuorinen A.,  2010, \mn@doi [Phys. Rev. D] {10.1103/PhysRevD.81.105021}, 81, 105021

\bibitem[\protect\citeauthoryear{Lavagno \& Pigato}{Lavagno \& Pigato}{2022}]{Lavagno:2022orw}
Lavagno A.,  Pigato D.,  2022, \mn@doi [Eur. Phys. J. A] {10.1140/epja/s10050-022-00885-6}, 58, 237

\bibitem[\protect\citeauthoryear{Letessier \& Rafelski}{Letessier \& Rafelski}{2002}]{Letessier:2002ony}
Letessier J.,  Rafelski J.,  2002, {Hadrons and Quark\textendash{}Gluon Plasma}.
Oxford University Press, \mn@doi{10.1017/9781009290753}

\bibitem[\protect\citeauthoryear{Liang \& Zhitnitsky}{Liang \& Zhitnitsky}{2016}]{Liang:2016tqc}
Liang X.,  Zhitnitsky A.,  2016, \mn@doi [Phys. Rev. D] {10.1103/PhysRevD.94.083502}, 94, 083502

\bibitem[\protect\citeauthoryear{Lugones \& Horvath}{Lugones \& Horvath}{2004}]{Lugones:2003un}
Lugones G.,  Horvath J.~E.,  2004, \mn@doi [Phys. Rev. D] {10.1103/PhysRevD.69.063509}, 69, 063509

\bibitem[\protect\citeauthoryear{Madsen}{Madsen}{1988}]{Madsen:1988zgf}
Madsen J.,  1988, \mn@doi [Phys. Rev. Lett.] {10.1103/PhysRevLett.61.2909}, 61, 2909

\bibitem[\protect\citeauthoryear{Madsen}{Madsen}{1998}]{Madsen1998}
Madsen J.,  1998, \mn@doi [Physical Review Letters] {10.1103/PhysRevLett.81.3311}, 81, 3311

\bibitem[\protect\citeauthoryear{Madsen}{Madsen}{1999a}]{Madsen1999}
Madsen J.,  1999a, \mn@doi [Physical Review Letters] {10.1103/PhysRevLett.85.10}, 85, 10

\bibitem[\protect\citeauthoryear{Madsen}{Madsen}{1999b}]{Madsen:1998uh}
Madsen J.,  1999b, \mn@doi [Lect. Notes Phys.] {10.1007/BFb0107314}, 516, 162

\bibitem[\protect\citeauthoryear{Madsen \& Riisager}{Madsen \& Riisager}{1985}]{Madsen:1985zx}
Madsen J.,  Riisager K.,  1985, \mn@doi [Phys. Lett. B] {10.1016/0370-2693(85)90956-6}, 158, 208

\bibitem[\protect\citeauthoryear{Madsen, Heiselberg  \& Riisager}{Madsen et~al.}{1986}]{Madsen:1986jg}
Madsen J.,  Heiselberg H.,   Riisager K.,  1986, \mn@doi [Phys. Rev. D] {10.1103/PhysRevD.34.2947}, 34, 2947

\bibitem[\protect\citeauthoryear{Martens, Greiner, Leupold  \& Mosel}{Martens et~al.}{2006}]{Martens:2006ac}
Martens G.,  Greiner C.,  Leupold S.,   Mosel U.,  2006, \mn@doi [Phys. Rev. D] {10.1103/PhysRevD.73.096004}, 73, 096004

\bibitem[\protect\citeauthoryear{Middleton, Zurm\"uhle, Klein  \& Kollarits}{Middleton et~al.}{1979}]{PhysRevLett.43.429}
Middleton R.,  Zurm\"uhle R.~W.,  Klein J.,   Kollarits R.~V.,  1979, \mn@doi [Phys. Rev. Lett.] {10.1103/PhysRevLett.43.429}, 43, 429

\bibitem[\protect\citeauthoryear{Mueller, Wang, Holt, Lu, O'Connor  \& Schiffer}{Mueller et~al.}{2004}]{PhysRevLett.92.022501}
Mueller P.,  Wang L.-B.,  Holt R.~J.,  Lu Z.-T.,  O'Connor T.~P.,   Schiffer J.~P.,  2004, \mn@doi [Phys. Rev. Lett.] {10.1103/PhysRevLett.92.022501}, 92, 022501

\bibitem[\protect\citeauthoryear{{Nakamura} \& {Frohlich}}{{Nakamura} \& {Frohlich}}{2006}]{2006LPI....37.1048N}
{Nakamura} Y.,  {Frohlich} C.,  2006, in {Mackwell} S.,  {Stansbery} E.,  eds, 37th Annual Lunar and Planetary Science Conference. Lunar and Planetary Science Conference.
p.~1048

\bibitem[\protect\citeauthoryear{Navarro, Frenk  \& White}{Navarro et~al.}{1997}]{Navarro:1996gj}
Navarro J.~F.,  Frenk C.~S.,   White S. D.~M.,  1997, \mn@doi [Astrophys. J.] {10.1086/304888}, 490, 493

\bibitem[\protect\citeauthoryear{Nesti \& Salucci}{Nesti \& Salucci}{2013}]{Nesti_2013}
Nesti F.,  Salucci P.,  2013, \mn@doi [Journal of Cosmology and Astroparticle Physics] {10.1088/1475-7516/2013/07/016}, 2013, 016–016

\bibitem[\protect\citeauthoryear{Niikura et~al.}{Niikura et~al.}{2019}]{Niikura:2017zjd}
Niikura H.,  et~al., 2019, \mn@doi [Nature Astron.] {10.1038/s41550-019-0723-1}, 3, 524

\bibitem[\protect\citeauthoryear{Pereira, Flores  \& Lugones}{Pereira et~al.}{2018}]{Pereira:2017rmp}
Pereira J.~P.,  Flores C.~V.,   Lugones G.,  2018, \mn@doi [Astrophys. J.] {10.3847/1538-4357/aabfbf}, 860, 12

\bibitem[\protect\citeauthoryear{Piotrowski, Ma\l{}ek, Mankiewicz, Soko\l{}owski, Wrochna, Zadro\ifmmode~\dot{z}\else \.{z}\fi{}ny  \& \ifmmode~\dot{Z}\else \.{Z}\fi{}arnecki}{Piotrowski et~al.}{2020}]{LECH___PhysRevLett.125.091101}
Piotrowski L.~W.,  Ma\l{}ek K.,  Mankiewicz L.,  Soko\l{}owski M.,  Wrochna G.,  Zadro\ifmmode~\dot{z}\else \.{z}\fi{}ny A.,   \ifmmode~\dot{Z}\else \.{Z}\fi{}arnecki A.~F.,  2020, \mn@doi [Phys. Rev. Lett.] {10.1103/PhysRevLett.125.091101}, 125, 091101

\bibitem[\protect\citeauthoryear{Pollack, Spergel  \& Steinhardt}{Pollack et~al.}{2015}]{Pollack:2014rja}
Pollack J.,  Spergel D.~N.,   Steinhardt P.~J.,  2015, \mn@doi [Astrophys. J.] {10.1088/0004-637X/804/2/131}, 804, 131

\bibitem[\protect\citeauthoryear{Price \& Salamon}{Price \& Salamon}{1986}]{Salamon:1986}
Price P.~B.,  Salamon M.~H.,  1986, \mn@doi [Phys. Rev. Lett.] {10.1103/PhysRevLett.56.1226}, 56, 1226

\bibitem[\protect\citeauthoryear{Price, Shirk, Osborne  \& Pinsky}{Price et~al.}{1978}]{PhysRevD.18.1382}
Price P.~B.,  Shirk E.~K.,  Osborne W.~Z.,   Pinsky L.~S.,  1978, \mn@doi [Phys. Rev. D] {10.1103/PhysRevD.18.1382}, 18, 1382

\bibitem[\protect\citeauthoryear{{Price} et~al.,}{{Price} et~al.}{1992b}]{1992Ap&SS.197..121P}
{Price} P.~B.,  et~al., 1992b, \mn@doi [\apss] {10.1007/BF00645077}, \href {https://ui.adsabs.harvard.edu/abs/1992Ap&SS.197..121P} {197, 121}

\bibitem[\protect\citeauthoryear{Price et~al.,}{Price et~al.}{1992a}]{Price1992}
Price P.~B.,  et~al., 1992a, \mn@doi [Astrophysics and Space Science] {10.1007/BF00645077}, 197, 121

\bibitem[\protect\citeauthoryear{Roessner, Ratti  \& Weise}{Roessner et~al.}{2007}]{Roessner:2006xn}
Roessner S.,  Ratti C.,   Weise W.,  2007, \mn@doi [Phys. Rev. D] {10.1103/PhysRevD.75.034007}, 75, 034007

\bibitem[\protect\citeauthoryear{Ruester, Werth, Buballa, Shovkovy  \& Rischke}{Ruester et~al.}{2006}]{Ruester:2005ib}
Ruester S.~B.,  Werth V.,  Buballa M.,  Shovkovy I.~A.,   Rischke D.~H.,  2006, \mn@doi [Phys. Rev. D] {10.1103/PhysRevD.73.034025}, 73, 034025

\bibitem[\protect\citeauthoryear{Saito, Hatano, Fukada  \& Oda}{Saito et~al.}{1990}]{PhysRevLett.65.2094}
Saito T.,  Hatano Y.,  Fukada Y.,   Oda H.,  1990, \mn@doi [Phys. Rev. Lett.] {10.1103/PhysRevLett.65.2094}, 65, 2094

\bibitem[\protect\citeauthoryear{Salucci}{Salucci}{2019}]{Salucci:2018hqu}
Salucci P.,  2019, \mn@doi [Astron. Astrophys. Rev.] {10.1007/s00159-018-0113-1}, 27, 2

\bibitem[\protect\citeauthoryear{Schaffner-Bielich}{Schaffner-Bielich}{2005}]{Schaffner-Bielich:2004lxd}
Schaffner-Bielich J.,  2005, \mn@doi [J. Phys. G] {10.1088/0954-3899/31/6/004}, 31, S651

\bibitem[\protect\citeauthoryear{Schäfer \& Wilczek}{Schäfer \& Wilczek}{1999}]{Sch_fer_1999}
Schäfer T.,  Wilczek F.,  1999, \mn@doi [Physical Review Letters] {10.1103/physrevlett.82.3956}, 82, 3956–3959

\bibitem[\protect\citeauthoryear{Shahrbaf, Blaschke, Typel, Farrar  \& Alvarez-Castillo}{Shahrbaf et~al.}{2022}]{Shahrbaf:2022upc}
Shahrbaf M.,  Blaschke D.,  Typel S.,  Farrar G.~R.,   Alvarez-Castillo D.~E.,  2022, \mn@doi [Phys. Rev. D] {10.1103/PhysRevD.105.103005}, 105, 103005

\bibitem[\protect\citeauthoryear{{Shirk} \& {Price}}{{Shirk} \& {Price}}{1978}]{1978ApJ...220..719S}
{Shirk} E.~K.,  {Price} P.~B.,  1978, \mn@doi [\apj] {10.1086/155955}, \href {https://ui.adsabs.harvard.edu/abs/1978ApJ...220..719S} {220, 719}

\bibitem[\protect\citeauthoryear{Singh~Sidhu \& Starkman}{Singh~Sidhu \& Starkman}{2020}]{SinghSidhu:2019tbr}
Singh~Sidhu J.,  Starkman G.~D.,  2020, \mn@doi [Phys. Rev. D] {10.1103/PhysRevD.101.083503}, 101, 083503

\bibitem[\protect\citeauthoryear{Song \& Coci}{Song \& Coci}{2022}]{song2022}
Song T.,  Coci G.,  2022, \mn@doi [Nuclear Physics A] {https://doi.org/10.1016/j.nuclphysa.2022.122539}, 1028, 122539

\bibitem[\protect\citeauthoryear{Spergel \& Steinhardt}{Spergel \& Steinhardt}{2000}]{Spergel:1999mh}
Spergel D.~N.,  Steinhardt P.~J.,  2000, \mn@doi [Phys. Rev. Lett.] {10.1103/PhysRevLett.84.3760}, 84, 3760

\bibitem[\protect\citeauthoryear{Stephanov}{Stephanov}{2004}]{Stephanov:2004wx}
Stephanov M.~A.,  2004, \mn@doi [Prog. Theor. Phys. Suppl.] {10.1143/PTPS.153.139}, 153, 139

\bibitem[\protect\citeauthoryear{Suwa, Yoshida, Shibata, Umeda  \& Takahashi}{Suwa et~al.}{2018}]{Suwa:2018uni}
Suwa Y.,  Yoshida T.,  Shibata M.,  Umeda H.,   Takahashi K.,  2018, \mn@doi [Mon. Not. Roy. Astron. Soc.] {10.1093/mnras/sty2460}, 481, 3305

\bibitem[\protect\citeauthoryear{Tisserand et~al.}{Tisserand et~al.}{2007}]{EROS-2:2006ryy}
Tisserand P.,  et~al., 2007, \mn@doi [Astron. Astrophys.] {10.1051/0004-6361:20066017}, 469, 387

\bibitem[\protect\citeauthoryear{Tran, Geller, Lehmann  \& Kaiser}{Tran et~al.}{2024}]{Tran:2023jci}
Tran T.~X.,  Geller S.~R.,  Lehmann B.~V.,   Kaiser D.~I.,  2024, \mn@doi [Phys. Rev. D] {10.1103/PhysRevD.110.063533}, 110, 063533

\bibitem[\protect\citeauthoryear{Turkevich, Wielgoz  \& Economou}{Turkevich et~al.}{1984}]{PhysRevD.30.1876}
Turkevich A.,  Wielgoz K.,   Economou T.~E.,  1984, \mn@doi [Phys. Rev. D] {10.1103/PhysRevD.30.1876}, 30, 1876

\bibitem[\protect\citeauthoryear{Vartanyan, Hajyan, Grigoryan  \& Sarkisyan}{Vartanyan et~al.}{2009}]{Vartanyan:2009zza}
Vartanyan Y.~L.,  Hajyan G.~S.,  Grigoryan A.~K.,   Sarkisyan T.~R.,  2009, \mn@doi [Astrophysics] {10.1007/s10511-009-9055-7}, 52, 300

\bibitem[\protect\citeauthoryear{Vartanyan, Hajyan, Grigoryan  \& Sarkisyan}{Vartanyan et~al.}{2012}]{Vartanyan:2012zz}
Vartanyan Y.~L.,  Hajyan G.~S.,  Grigoryan A.~K.,   Sarkisyan T.~R.,  2012, \mn@doi [Astrophysics] {10.1007/s10511-012-9216-y}, 55, 98

\bibitem[\protect\citeauthoryear{Vovchenko, Brandt, Cuteri, Endr\H{o}di, Hajkarim  \& Schaffner-Bielich}{Vovchenko et~al.}{2021}]{Vovchenko:2020crk}
Vovchenko V.,  Brandt B.~B.,  Cuteri F.,  Endr\H{o}di G.,  Hajkarim F.,   Schaffner-Bielich J.,  2021, \mn@doi [Phys. Rev. Lett.] {10.1103/PhysRevLett.126.012701}, 126, 012701

\bibitem[\protect\citeauthoryear{Watkins, van~der Marel, Sohn  \& Wyn~Evans}{Watkins et~al.}{2019}]{Watkins_2019}
Watkins L.~L.,  van~der Marel R.~P.,  Sohn S.~T.,   Wyn~Evans N.,  2019, \mn@doi [The Astrophysical Journal] {10.3847/1538-4357/ab089f}, 873, 118

\bibitem[\protect\citeauthoryear{Watts \& Reddy}{Watts \& Reddy}{2007}]{Watts:2006hk}
Watts A.~L.,  Reddy S.,  2007, \mn@doi [Mon. Not. Roy. Astron. Soc.] {10.1111/j.1745-3933.2007.00336.x}, 379, L63

\bibitem[\protect\citeauthoryear{Weber}{Weber}{2005}]{Weber:2004kj}
Weber F.,  2005, \mn@doi [Prog. Part. Nucl. Phys.] {10.1016/j.ppnp.2004.07.001}, 54, 193

\bibitem[\protect\citeauthoryear{Wiktorowicz, Drago, Pagliara  \& Popov}{Wiktorowicz et~al.}{2017}]{Wiktorowicz:2017swq}
Wiktorowicz G.,  Drago A.,  Pagliara G.,   Popov S.~B.,  2017, \mn@doi [Astrophys. J.] {10.3847/1538-4357/aa8629}, 846, 163

\bibitem[\protect\citeauthoryear{Williams, Turyshev  \& Boggs}{Williams et~al.}{2004}]{Williams:2004qba}
Williams J.~G.,  Turyshev S.~G.,   Boggs D.~H.,  2004, \mn@doi [Phys. Rev. Lett.] {10.1103/PhysRevLett.93.261101}, 93, 261101

\bibitem[\protect\citeauthoryear{Witten}{Witten}{1984}]{Witten:1984rs}
Witten E.,  1984, \mn@doi [Phys. Rev. D] {10.1103/PhysRevD.30.272}, 30, 272

\bibitem[\protect\citeauthoryear{Zhitnitsky}{Zhitnitsky}{2003}]{Zhitnitsky:2002qa}
Zhitnitsky A.~R.,  2003, \mn@doi [JCAP] {10.1088/1475-7516/2003/10/010}, 10, 010

\bibitem[\protect\citeauthoryear{de Carvalho, Malheiro, Carlson, Frederico, Fiolhais, Scoccola  \& Grunfeld}{de~Carvalho et~al.}{2010}]{deCarvalho:2010wdj}
de Carvalho S.~M.,  Malheiro M.,  Carlson B.~V.,  Frederico T.,  Fiolhais M.,  Scoccola N.,   Grunfeld A.~G.,  2010, \mn@doi [Nucl. Phys. B Proc. Suppl.] {10.1016/j.nuclphysbps.2010.02.049}, 199, 308

\makeatother
\end{thebibliography}

\appendix

\section{Exploration of the parameter space}\label{paramspace}

\begin{figure}
\begin{centering}
\includegraphics[width=\columnwidth]{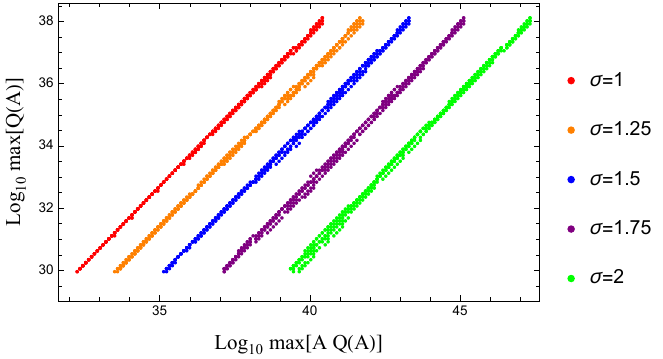}
    \caption{Correlation between the maximum mass and the maximum size of the final distribution, for different values of $\sigma$ and for $\text{log}_{10}\beta$ varying from -5 to -6. }
    \label{dimensionemassa}
    \end{centering}
\end{figure}

\begin{figure}
\begin{centering}
\includegraphics[width=\columnwidth]{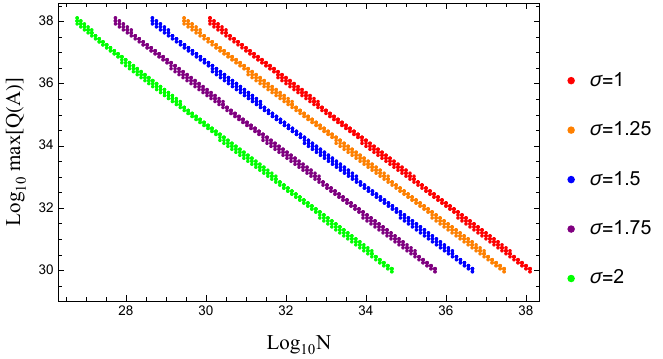}
    \caption{Correlation between the final number of strangelets and the maximum size of the final distribution, for different values of $\sigma$ and for $\text{log}_{10}\beta$ varying from -5 to -6. }
    \label{dimensionenumero}
    \end{centering}
\end{figure}

\begin{figure}
\begin{centering}
\includegraphics[width=\columnwidth]{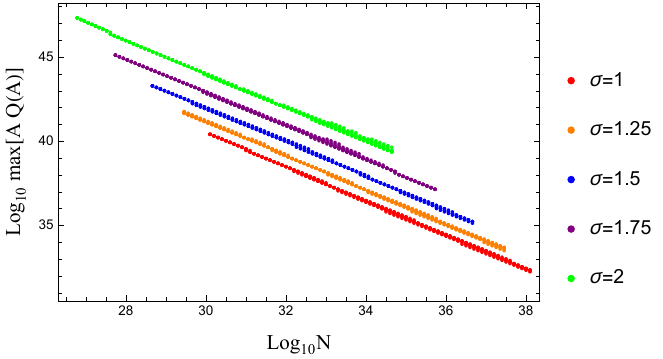}
    \caption{Correlation between the number of strangelets and the maximum mass, for different values of $\beta$ and for $\sigma$ varying from 1 to 2. }
    \label{massanumero}
    \end{centering}
\end{figure}

\begin{figure}
\begin{centering}
\includegraphics[width=\columnwidth]{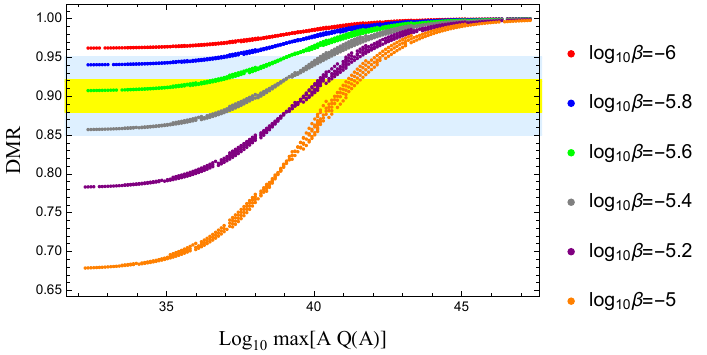}
    \caption{Correlation between the maximum mass and the DMR, varying $\beta$ for $T_\mathrm{evap}=150$ MeV}
    \label{dmrmassa}
    \end{centering}
\end{figure}

\begin{figure}
\begin{centering}
\includegraphics[width=\columnwidth]{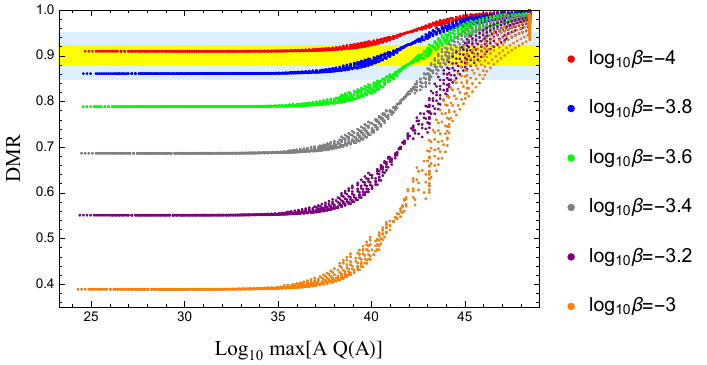}
    \caption{Correlation between the maximum mass and the DMR, varying $\beta$ for $T_\mathrm{evap}=50$ MeV}
    \label{dmrmassa2}
    \end{centering}
\end{figure}

In order to produce the distributions in \autoref{distr} we systematically explored the parameter space.
To generate the series of plots presented in \autoref{dimensionemassa}, \autoref{dimensionenumero}, \autoref{massanumero}, \autoref{dmrmassa}, and \autoref{dmrmassa2}, we employed a detailed evaporation routine. This routine was executed over a range of parameters: the initial central value of the distribution $\mu_{\text{max}} \in \left[30,38\right]$, the standard deviation $\sigma \in \left[1,2\right]$, and $\text{log}_{10}\beta \in \left[-5,-6\right]$ for $T_\mathrm{evap}=150$ MeV. Moreover, we used the range $\text{log}_{10}\beta \in \left[-3,-4\right]$ for $T_\mathrm{evap}=50$ MeV, to show a strong correlation between the $\beta$ parameter and the evaporation temperature $T_\mathrm{evap}$ (see \autoref{sec:evaporation}).

The selection of these parameter ranges was informed by preliminary investigations, which explored a broader parameter space. However, simulations conducted outside those ranges failed to satisfy all the constraints delineated in \autoref{obsconstr}. Consequently, to ensure the relevance with respect to the established observational constraints, our analysis was restricted to these specific ranges.

\autoref{dimensionemassa} shows a linear correlation between the log of the maximum of the mass distribution $AQ(A)$, LMAQ, and the log of the maximum of the number distribution $Q(A)$, LMQ. This correlation is observed when the parameters $\sigma$ (of the pre-evaporation distribution $P(A)$) and $\beta$ (from \autoref{evapobeta}) are held constant. 

A similar linear correlation exists, for fixed values of $\sigma$ and $\beta$, between the log of the total number of strangelets after evaporation and LMQ (\autoref{dimensionenumero}) or LMAQ (\autoref{massanumero}).

A non-linear correlation is identified instead in \autoref{dmrmassa} and \autoref{dmrmassa2}, for various values of $\log_{10}\beta$, between the dark matter ratio (DMR) and LMAQ. This relationship suggests that, to achieve the correct DMR, we have to significantly suppress the evaporation rate. It is important to note that, by choosing a lower value of $T_\mathrm{evap}$, the suppression of the evaporation rate can be less dramatic. In other words, the role played by $\beta$ and $T_\mathrm{evap}$ is very similar.



\bsp	
\label{lastpage}
\end{document}